\documentclass[a4paper,11pt]{article}
\pdfoutput=1 
\usepackage{jcappub}

\usepackage[T1]{fontenc} 
\usepackage{placeins}
\usepackage{rotating}

\usepackage{mathtools}

\usepackage{textcomp}

\usepackage{array, caption, floatrow, tabularx, makecell, booktabs}%
\usepackage{ulem}
\usepackage{amsfonts}
\usepackage{amssymb}
\usepackage{amsthm}
\usepackage{amsbsy}
\usepackage{bm}
\usepackage{amsmath}

\usepackage{mathrsfs}
\usepackage{slashed}
\usepackage{graphicx}
\usepackage{tikz-cd}
\usepackage{fancyhdr}
\usepackage{tikz}
\usetikzlibrary{arrows.meta,
                tikzmark}

\usepackage{comment}
\usepackage{xcolor}
\usepackage{capt-of}
\usepackage{url}
\usepackage{multirow}
\usepackage{chngcntr}
\usepackage[font=footnotesize]{caption}

\newcommand{\beq}{\begin{equation}}
\newcommand{\eeq}{\end{equation}}
\newcommand{\bea}{\begin{eqnarray}}
\newcommand{\eea}{\end{eqnarray}}
\newcommand{\Trh}{T_{\rm RH}}
\newcommand{\Tmax}{T_{\rm max}}
\newcommand{\arh}{a_{\rm RH}}
\newcommand{\aend}{a_{\rm end}}

\title{Leptogenesis and Low Reheating Temperatures}

\author[a]{Marcos A. G. Garcia,}
\author[b,1]{Stephen E. Henrich,\note{Corresponding author.}}
\author[b]{Wenqi Ke,}
\author[b]{ and Keith A. Olive}

\affiliation[a]{Departamento de F\'isica Te\'orica, Instituto de F\'isica, Universidad Nacional Aut\'onoma de M\'exico, Ciudad de M\'exico C.P. 04510, Mexico} 
\affiliation[b]{William I.~Fine Theoretical Physics Institute, School of Physics and Astronomy, University of Minnesota, Minneapolis, MN 55455, USA}

\emailAdd{marcos.garcia@fisica.unam.mx}
\emailAdd{henri455@umn.edu}
\emailAdd{wke@umn.edu} 
\emailAdd{olive@umn.edu}

\abstract{We study leptogenesis during non-instantaneous reheating in the canonical type-I seesaw framework, with the dominant source of right-handed neutrino (RHN) production being non-thermal from inflaton decays ($\phi \rightarrow NN$). While matter-like reheating ($w_\phi=0$) fails to be compatible with standard leptogenesis for very low reheating temperatures, the situation is strikingly different for generalized Starobinsky potentials approximated by $V(\phi)\propto\phi^k$ with $k\geq4$ about the minimum. In the latter cases,  the observed baryon asymmetry can readily be obtained for arbitrarily low reheating temperatures above the BBN bound of $\sim4$~MeV. We study radiation-like reheating ($w_\phi=1/3$, $k=4$) in detail, showing that the evolving effective mass of the inflaton condensate leads to kinematic shutoff of the $\phi\rightarrow NN$ channel, which qualitatively changes the leptogenesis dynamics. We include a detailed treatment of the effects of fragmentation of the inflaton condensate. The final baryon asymmetry depends primarily on only two parameters: the inflaton-RHN coupling, $y_{\phi NN}$, and the CP-violating parameter $|\epsilon|$. Interestingly, the final asymmetry is largely insensitive to the RHN mass, the reheating temperature, and the RHN decay rate. While we focus on fermionic reheating, we show that the general features of these results also hold for bosonic reheating to scalars.
}

\begin{document}

\begin{flushright}
UMN--TH--4534/26, FTPI--MINN--26/14   \\
July 2026
\end{flushright}

\maketitle
\flushbottom

\section{Introduction}
\label{intro}

There are a number of elements central to early Universe cosmology: for example, big bang nucleosynthesis (BBN)  \cite{CFOY,coc18,foyy}, baryogenesis/leptogenesis \cite{Kolb:1983ni,Fukugita:1986hr,Olive:1994xw,Riotto:1999yt,Bodeker:2020ghk}, and inflation \cite{Staro,inflation,reviews}. Inflation resets the initial conditions for Standard Big Bang cosmology, and through the process of reheating, facilitates a transition to the period of early radiation domination necessary for successful BBN. BBN however, only requires rather minimal reheating, with a temperature, $\Trh > 4$~MeV \cite{tr4}. This allows for cosmological histories with low reheating temperatures, which have been a subject of much recent interest for instance in dark matter model building \cite{Bhattiprolu:2022sdd, Cosme:2023xpa, Silva-Malpartida:2023yks, Arcadi:2024wwg, Boddy:2024vgt, Belanger:2024yoj, Amiri:2025ras, Henrich:2025pca, Henrich:2026tox}. 
Ignoring the details of the reheating process, such a low reheating temperature may be problematic for generating a baryon asymmetry. For example, naively, one may argue that
any mechanism which relies on interactions mediated by $(B+L)$-violating electroweak sphaleron processes ~\cite{Kuzmin:1985mm,Arnold:1987mh,spha2} requires temperatures above the electroweak scale of order 100 GeV. 
Indeed, while there are many viable mechanisms for generating the baryon asymmetry, among the most economic is leptogenesis \cite{Fukugita:1986hr}. A minimum of two right-handed neutrinos (RHNs), $N_i$, are required beyond the Standard Model field content. These two states would suffice to generate the CP violation in the decay of the lighter state, $N_1$, producing a lepton asymmetry and at the same time provide masses via the see-saw mechanism \cite{seesaw} to two of the three Standard Model (SM) neutrinos (all that is required by neutrino oscillation experiments). The lepton asymmetry is then partially transformed to a baryon asymmetry via sphaleron interactions. 

The origin of the right-handed states may be thermal 
\cite{Buchmuller:2002rq,Buchmuller:2003gz,Chankowski:2003rr,Giudice:2003jh}. However, the lower bound on the right-handed mass in thermal leptogenesis where the right-handed states are produced from the thermal bath after reheating implies a bound on the reheating temperature of $\Trh \gtrsim 10^{10}$~GeV \cite{DIbound}. This does not correspond to a low reheating temperature. Alternatively, right-handed neutrinos may be produced directly from inflaton decay
\cite{Lazarides:1990huy,Campbell:1992hd,Giudice:1999fb,Asaka:1999yd,Hahn-Woernle:2008tsk}. This is inherently a non-thermal process for $M_N > \Trh$ if we assume instantaneous reheating, and the right-handed neutrinos are always out of equilibrium. 
We must only constrain $M_N < m_\phi/2$, where $m_\phi$ is the inflaton mass.\footnote{The production of RHN with $M_N > m_\phi$ may be possible during preheating \cite{Kanemura:2025rct}.} As in the case of thermal leptogenesis, right-handed neutrino decay produces a lepton asymmetry which must be converted to a baryon asymmetry and therefore requires sphaleron interactions to be in equilibrium. If reheating is an instantaneous process (which it is not), we would therefore require that $\Trh > T_{\rm sph} \sim 130$~GeV \cite{DOnofrio:2014rug}. More precisely, we must only require that the sphalerons were in equilibrium at some point in the early universe together with a non-zero lepton asymmetry (or more precisely a non-zero $B-L$ asymmetry). This is possible even if $\Trh < T_{\rm sph}$ when the dynamics of the reheating process are taken into account. For related studies of leptogenesis during reheating see \cite{Hahn-Woernle:2008tsk,Hamada:2015xva,Zhang:2023oyo,Barman:2024ujh}, though these works were not focusing on low reheating temperatures. 

While the specific details of reheating will depend on the model of inflation, there are many generic features of reheating which we can utilize to draw important conclusions about leptogenesis. Here, we will only consider models of inflation where the vacuum energy during inflation is set by the amplitude of the cosmic microwave background (CMB) anisotropies. This typically results in a Hubble parameter during inflation of order $10^{13}$~GeV. We will also assume that the primary mechanism for reheating occurs through the decay of the inflaton to either fermions or bosons in the final state. 
An example of such a model (though we do not tie ourselves to this model) is the Starobinsky model \cite{Staro} of inflation, which is described by the scalar potential
\beq
V(\phi) = \frac{3}{4} \lambda  M_{P}^{4}\left(1-e^{-\sqrt{\frac{2}{3}} \frac{\phi}{M_{P}}}\right)^{2} \, ,
\label{Staro}
\eeq
where $M_P \simeq 2.4 \times 10^{18}$~GeV is the reduced Planck mass. 
For this potential, the Hubble parameter for $\phi \gg M_P$ is $H = \frac12 \sqrt{\lambda} M_P$ and the inflaton mass is $m_\phi = \sqrt{\lambda} M_P$. The coefficient $\lambda$ is fixed by \cite{GKMO2,Ellis:2021kad}
\begin{equation}
\label{eq:infnorm}
\lambda \; \simeq \; \frac{24\pi^2 A_{s}}{N_*^2} \, , 
\end{equation}
where $A_s$ is the amplitude of scalar perturbations and is observationally constrained to $\ln(10^{10} A_s) = 3.044 \pm 0.014$ at 68\% CL from \textit{Planck} 2018 TT,TE,EE+lowE+lensing data~\cite{Planck}, corresponding to $A_s \simeq 2.10 \times 10^{-9}$. 
$N_*$ is the number of $e$-folds from horizon exit of the pivot scale to the end of inflation and depends on the reheating temperature \cite{LiddleLeach,Martin:2010kz, Ellis:2021kad,Ellis:2025zrf}. An approximate expression for $N_*$ in this model is \cite{Ellis:2025zrf}
$N_* + \frac13\ln N_* = 59.55 + \frac13\ln (\Trh/M_P)$.
Thus for a given value of $\Trh$, $N_*$ is determined and that fixes the value of $\phi_*$ (the value of $\phi$ at the pivot scale) which in turn determines the slow-roll parameters and the CMB observables such as the tilt of the scalar anisotropy spectrum, $n_s$ and the tensor-to-scalar ratio, $r$. For a high reheating temperature, $\Trh = 10^{14}$~GeV, $N_* \simeq 55$, and $n_s \simeq 0.965$ 
which can be compared with the Planck value of $n_s = 0.9649 \pm 0.0044$ \cite{Planck}. In this case, $\lambda \simeq 1.5 \times 10^{-10}$ and $m_\phi \simeq 3.0 \times 10^{13}$~GeV.  In contrast, for a reheating temperature $\Trh = 1$~GeV, $N_* \simeq 44$, and $n_s \simeq 0.956$.
In this case, $\lambda \simeq 2.4 \times 10^{-10}$ and $m_\phi \simeq 3.7 \times 10^{13}$~GeV. 

For the potential in Eq.~(\ref{Staro}),
accelerated expansion ends when $\phi = \phi_{\rm end} \approx 0.63 M_P$  and when the cosmological scale factor is $a = a_{\rm end}$. Subsequently the inflaton condensate begins a series of oscillations about its minimum at $\phi = 0$. During these oscillations the energy density of the inflaton redshifts as $\rho_\phi \propto a^{-3}$ and inflaton decays quickly establish a thermal bath with a maximum temperature\footnote{Note that independent of $\Gamma_{\phi ff}$, there is a ``minimal'' maximum temperature of order $10^{12}$~GeV for radiation produced by inflaton scattering mediated by gravity \cite{Clery:2021bwz}. Similar processes  will produce RHNs as discussed in Section \ref{gravprod}. } \cite{Giudice:2000ex, Ellis:2015jpg} at $a = a_{\rm max} = (8/3)^{2/5}a_{\rm end}$ and 
\beq
\alpha_{\rm max} T_{\rm max}^4 =  \frac{\sqrt{3}}{4} \left(\frac38 \right)^\frac35 \Gamma_{\phi \to ff} \rho_{\rm end}^\frac12 M_P \, ,
\eeq
where $\alpha_{\rm max} = g_*(a_{\rm max}) \pi^2/30$ and $\Gamma_{\phi \to ff} = y_{\phi ff}^2 m_\phi/8\pi$ is the inflaton decay rate to fermions which couple to the Standard Model. The inflaton energy density at the end of inflation is $\rho_{\rm end} = \frac32 V(\phi_{\rm end})$, 
$g_*(a)$ is the number of relativistic degrees of freedom at $a$, and $y_{\phi ff}$ is the coupling of the inflaton to its fermion decay products.\footnote{Indeed in the Standard Model, there are no fermions which can couple to the inflaton in this way. The inflaton can couple to a pair of RHNs, but we treat this decay rate separately for leptogenesis rather than for reheating. The minimal supersymmetric Standard Model (MSSM) offers the possibility for inflaton decays to a pair of Higgsinos. As such, we will take $g_* = 915/4$ at high temperatures when all MSSM degrees of freedom are in the thermal bath. } As inflaton decays proceed, the temperature of the thermal bath decreases as $T \propto a^{-3/8}$ \cite{Scherrer:1984fd,Giudice:2000ex,GKMO1,GKMO2} until reheating is achieved (defined by $\rho_\phi (\arh) = \rho_{\rm R}(\arh)$). The coupling $y_{\phi ff}$ then determines the reheating temperature
\beq
\alpha_{\rm RH} \Trh^4 = \frac{12}{25} \Gamma_{\phi\to ff}^2 M_P^2 \, ,
\eeq
where now $\alpha_{\rm RH} = g_*(\arh) \pi^2/30$.

Let us now suppose that in addition to the decays which lead to the thermal bath (controlled by the coupling $y_{\phi ff}$), the inflaton also decays to right-handed neutrinos with a branching fraction $Br = (\Gamma_{\phi \rightarrow NN}/\Gamma_{\phi \rm tot})$, where $\Gamma_{\phi\rightarrow NN}$ is the rate of inflaton decay to right-handed neutrinos. For $Br \ll 1$, $\Gamma_{\phi \to ff} \approx \Gamma_{\phi \rm tot}$.  We further assume that $m_\phi/2 > M_N > \Tmax$. The condition that $m_\phi/2> \Tmax$ requires only that $\Trh < 2.0 \times 10^{11}$~GeV.  Integrating the Boltzmann equation for the production of $N$ (to be discussed in more detail in Section \ref{sec:fermion}), one sees that right-handed neutrinos continue to be produced non-thermally via inflaton decays throughout the entirety of reheating, that is for all scale factors $a_{\rm end}<a<a_{\rm RH}$. 

We expect right-handed neutrinos to decay through the coupling to left-handed leptons and the Higgs bosons, $y_N H L N$. 
These decays will occur approximately when $\Gamma_N \propto y_N^2 \sim H$, and we label the scale factor at this time by $a_N$.
Furthermore, for large $M_N$ (as considered here) and so long as 
\beq
y_N \gtrsim 1.7\times 10^{-19} \left(\frac{10^{12} \text{ GeV}}{M_N}\right)^{1/2}\left(\frac{10^{12} \text{ GeV}}{T_{\rm max}}\right)^2 \, ,
\eeq
or
\beq
y_N \gtrsim 1.5\times10^{-10}\left(\frac{10^{12} \text{ GeV}}{M_N}\right)^{1/2}\left(\frac{1 \text{ GeV}}{T_{\rm RH}}\right) \, ,
\eeq
any right-handed neutrino produced at $T > T_{\rm sph}$ will decay prior to sphaleron decoupling and produce a lepton asymmetry, $n_L = n_{B-L}$ proportional to $\epsilon n_N$, where $\epsilon$ is a measure of the CP violation in N-decay.

For any right-handed neutrino decay at $T>T_{\rm sph}$, 
sphaleron transitions are in equilibrium.
 In equilibrium, the baryon density is given by the $B-L$ asymmetry with \cite{HT}
\beq
n_B = \frac{8N_F + 4N_H}{22 N_F + 13 N_H} n_{B-L} \, ,
\label{coef}
\eeq
where $N_F$ is the number of fermion generations and $N_H$ is the number of Higgs doublets. In the Standard Model, we have $N_F = 3$, $N_H = 1$, such that the coefficient relating $n_B$ and $n_{B-L}$ is 28/79. In the MSSM, $N_H = 2$, and the ratio of $B$ to $B-L$ is 8/23. We will use the MSSM value of 8/23 throughout this work for concreteness, although it is important to note that none of our key results will rely on supersymmetry. In leptogenesis, where purely a lepton asymmetry is generated, $B-L = -L$. 
The baryon to entropy ratio evaluated at $T_{\rm sph}$ for sufficiently high temperature reheating where $N$ decays occur approximately at the same time as reheating is complete is given by 
\beq
Y_B|_{T=T_{\rm sph}}\equiv \frac{n_B}{s}|_{T=T_{\rm sph}} \sim \frac{8}{23} \epsilon \frac{n_N}{s} \sim \frac{8}{23} \epsilon \frac{n_N}{n_{\rm R}} \sim \frac{8}{23} \epsilon Br  \, ,
\label{Eq:instRehApprox}
\eeq
up to numerical factors relating  the entropy density and number density of radiation, $n_{\rm R}$ \footnote{In fact, thermalization may significantly alter the estimate of the asymmetry in Eq.~(\ref{Eq:instRehApprox}) because the naive radiation number density obtained from the branching ratio will not be the same as the radiation number density after thermalization. We examine this in detail in Section \ref{k2}.}.

 This asymmetry can easily be large enough to account for the observed baryon asymmetry of the Universe. The most recent determination of the baryon density from CMB measurements by Planck \cite{Planck}
and BBN \cite{Yeh:2026pil} gives $n_B/n_\gamma = 6.12 \times 10^{-10}$. In standard big bang cosmology, $s/n_\gamma = 7.04$, and thus
we have
\beq
Y_B|_{\rm obs} = 8.69 \times 10^{-11} \, .
\label{YBobs}
\eeq
However, if $\Trh < T_{\rm sph}$, radiation and right handed neutrinos continue to be produced via inflaton decays at temperatures below the electroweak scale. The right handed neutrinos will continue to decay, adding to the lepton asymmetry, but this asymmetry will not be converted to a baryon asymmetry as sphaleron transitions are no longer in equilibrium. At $T = \Trh$, the baryon asymmetry will therefore be diluted relative to the asymmetry at $T=T_{\rm sph}$ \cite{Davidson:2000dw} so that 
\beq
\frac{n_B}{s}|_{T=\Trh} \sim \frac{8}{23} \epsilon Br \left(\frac{\Trh}{T_{\rm sph}}\right)^\gamma \, .
\label{dilute}
\eeq
For a typical inflationary model as discussed above (where the equation of state of the oscillating inflaton condensate is $w_\phi=0$), 
the entropy density and baryon density will scale differently at $T < T_{\rm sph}$,
with
\beq
s \propto T^3 \qquad {\rm and} \qquad n_B \propto a^{-3} \propto T^8 \, .
\eeq
Therefore the dilution factor in Eq.~(\ref{dilute}) is given by $\gamma = 5$ for $w_\phi=0$.
Thus for $T \lesssim 1$~GeV, this corresponds to a dilution factor $> 10^{10}$, and would be far too small to account to the observed asymmetry.
In fact, as we will see below, there are additional factors which dilute the baryon asymmetry when non-instantaneous reheating is taken into account and $\Trh < m_\phi$. This will be discussed in more detail in Section \ref{k2}.

Recall that to generate a lepton asymmetry, we need at least two RHNs with CP violation present in the interference between the tree-level and one-loop decays of the lightest RHN, designated as $N_1$.  More generally we may expect three massive RHNs with masses $M_i$. 
We associate the mass of the lightest state, $M_1$ with $M_N$ and following \cite{kmov}, we assume $2 M_N \ll m_\phi \ll M_{2,3}$. 
The CP asymmetry parameter $\epsilon$ can be determined by \cite{luty,CPviol}
\begin{equation}
    \epsilon\simeq -\frac{3\delta_{\rm eff}|y_3^2|}{16\pi }\frac{M_1}{M_3} \, ,
    \label{Eq:epsilon}
\end{equation}
where $y_3$ is the largest effective Yukawa coupling and is associated with the heaviest (3rd generation for a normal hierarchy) neutrino.  
The CP violating phase $\delta_{\rm eff}$ 
is given by
\begin{equation}
    \delta_{\rm eff} = \frac{1}{|y_3|^2} \frac{{\rm Im}(yy^\dagger)^2_{13}}{(yy^\dagger)_{11}} \, .
\end{equation}
  This leads to an expression for the baryonic asymmetry\footnote{Here we have used the relation $m_{\nu_i} = y_i^2 v^2 \sin^2 \beta / M_i$ where $v = 174$~GeV and $\tan \beta = 1$, being the ratio of the two MSSM Higgs expectation values. Note that the survival of the lepton (and baryon) asymmetry requires that effective operators such as the $\Delta L = 2$ operator $\frac{y_i^2}{M_i} LLHH$ are out of equilibrium at temperatures between $T_{\rm sph}$ and $\Tmax$. This places a limit on $\frac{y_i^2}{M_i} \lesssim 10^{-14}$ or equivalently an upper limit on neutrino masses $m_{\nu_i} \lesssim 0.3$~eV \cite{Fukugita:1986hr,Olive:1994xw,HT,Campbell:1990fa,Campbell:1991at,Fischler:1990gn,Ibanez:1992aj}.  } \cite{kmov} :
\begin{equation}
    Y_B\simeq 7\times 10^{-5}\delta_{\rm eff}\frac{n_{N_1}}{s}\left(\frac{m_{\nu_i}}{0.05{\rm eV}}\right)\left(\frac{M_1}{10^{12}{\text{ GeV}}}\right) \, ,
    \label{analyyb}
\end{equation}
where $i = 2,3$.
We assume $M_1=M_N$ is the lightest RHN mass, and moving forward we will simply use $M_N$.

In this work we explore a simple modification of the reheating process which will allow successful leptogenesis at significantly lower reheating temperatures, even as low as the BBN limit of $\Trh > 4$~MeV \cite{tr4}. Nevertheless, throughout, we assume a high energy scale for inflation, with $H\sim \mathcal{O}(10^{13})$~GeV. 
In what follows, we first consider a generalization of the Starobinsky potential which allows for successful inflation, but alters the dynamics of the reheating process, including the effects of inflaton fragmentation. In Section \ref{sec:fermion}, we present the Boltzmann equations used for our numerical results and our analytical approximations. 
We return to a standard (though non-instantaneous) matter-like reheating scenario in Section \ref{k2}, where we fully demonstrate the difficulty for leptogenesis with low reheating temperatures. In Section \ref{k4}, we show how this problem is alleviated if
the minimum of inflaton potential is quartic rather than quadratic. 
In this section we provide our calculation of the baryon asymmetry along with numerical results for the baryon asymmetry and the viable parameter space. We comment on the gravitational production of RHNs and radiation in Section \ref{gravprod}. In Section \ref{other} we consider additional generalizations of the Starobinsky potential as well as scenarios with inflaton decays to scalar final states. Our conclusions are given in Section \ref{summary}.

\section{Generalized Starobinsky-like potentials}

Instead of Eq.~(\ref{Staro}), 
in this section we consider 
\beq
V(\phi) = \frac{3}{4} \lambda  M_{P}^{4}\left(1-e^{-\sqrt{\frac{2}{3}} \frac{\phi}{M_{P}}}\right)^{k} \, ,
\label{Starok}
\eeq
for even $k$. The Starobinsky model described in Section \ref{intro} corresponds to $k=2$. As is the case for the Starobinsky model, models with $k\ne2$ can also be easily derived from no-scale supergravity \cite{Ellis:2025zrf}.  During inflation (at large field values), the potential resembles that of the Starobinsky model and exhibits a flat plateau. For $k=4$, exponential expansion ends at
$\phi_{\rm end} \simeq 1.16 M_P$ and 
the temperature dependence in $N_*$ drops out \cite{GKMO1}, so that $N_* = 55.8$ for any $\Trh$. In this case, $\phi_* \simeq 6.18 M_P$, and $\lambda$ is essentially unchanged so that $\lambda = 1.5 \times 10^{-10}$ as it is for $k=2$ and $\rho_{\rm end}^\frac14 = (\frac32 V(\phi_{\rm end}))^\frac14 = 5.4 \times 10^{15} $~GeV.  The spectral tilt is also almost unchanged and $n_s = 0.965$ in this case as well.

Around the minimum at $\phi = 0$, the potential can be approximated by
\beq
V(\phi) \approx \frac{2^{\frac{k}{2}-2}}{3^{\frac{k}{2}-1}} \lambda \phi^k M_P^{4-k}  \, .
\label{appk}
\eeq
During the period of oscillations after inflation, the equation of state in general is given by $w_\phi = (k-2)/(k+2)$ and thus
$w_\phi =1/3$ for $k=4$. That is, these oscillations behave like radiation in contrast to the matter-like oscillations with $k=2$. Though the inflaton in its vacuum state is massless, the inflaton decays through its effective mass when $\phi \ne 0$. Its decays to SM radiation produce a maximum temperature at $a= a_{\rm max} = (4/3) \aend$, where
\beq
\alpha_{\rm max} \Tmax^4 = \left( \frac{ 3^\frac{11}{4}}{256 \pi} \right) \lambda^\frac14 \rho_{\rm end}^\frac34 M_P y_{\phi ff}^2 \, .
\label{tmax4}
\eeq

However, the reheating process for $k\ge4$
differs sharply from that of $k=2$ when the inflaton decay products are dominated by fermions in the final state. Specifically, self interactions of the inflaton can lead to the fragmentation of the inflaton condensate, taking energy out of the condensate creating a bath of inflaton quanta (which are of very low mass proportional to the density of the condensate left unfragmented) \cite{Garcia:2023eol,Garcia:2023dyf}. If reheating is not complete before fragmentation occurs, the reheating process may be severely affected. 

Fragmentation begins when the rate for converting energy in the condensate to particle quanta becomes comparable to the expansion rate. For $k=4$, this occurs when the scale
factor $a = a_\beta \simeq 90 \aend$ \cite{Garcia:2023dyf}.\footnote{Note that this value of $a_\beta$ has been calculated here for the Starobinsky-like potential in Eq.~(\ref{Starok}). See Appendix \ref{sec:frag} for details. In \cite{Garcia:2023dyf}, $a_\beta \simeq 180 \aend$, valid for the T-models of inflation with $k=4$. They differ due to the higher value of the coupling $\lambda$ used here.} For $a < a_\beta$, the effects of fragmentation can be ignored. 
In this case, the Boltzmann equation of the inflaton energy density (discussed more fully below) can be easily solved assuming $\Gamma_{\phi \to ff} \ll  H$ at early times, which gives us the well-known scaling $\rho_\phi\propto a^{-6k/(k+2)}\propto a^{-4}$ for $k=4$. We define $\xi \equiv \rho_{\bar{\phi}}/\rho_{\delta \phi}$ as the ratio of the energy density remaining in the condensate to that in particles. After a sudden drop in $\xi$ at $a = a_\beta$, $\xi$ continues to drop as $\xi = \xi_0 (a_\beta/a)^b$.
From numerical results, $\xi_0 \simeq 1.1$ and $b \simeq 1.3$ \cite{Garcia:2023dyf}.
When $a>a_\beta$:\begin{equation}\begin{aligned}
     & \rho_{\bar{\phi}}(a)\simeq \xi _0\rho_{\bar{\phi}}(a_\beta)\left(\frac{a_\beta}{a}\right)^{b+4}=\xi_0\rho_{\rm end }\left(\frac{a_{\rm end}}{a_\beta}\right)^4\left(\frac{a_\beta}{a}\right)^{b+4}\equiv \tilde{\rho}_{\bar{\phi}} /a^{b+4} \, , \\& \rho_{\delta \phi}(a)=  \rho_{\rm end} \left(\frac{a_{\rm end}}{a }\right)^4\equiv \tilde{\rho}_{\delta \phi } /a^{4} \, .
      \end{aligned}
\end{equation} 
For further details on the evolution of $\rho_{\bar \phi}$, $\rho_{\delta \phi}$, and $\xi$, see Fig.~\ref{frag} and the related discussion in Appendix \ref{sec:frag}.

Although for $k=4$, the inflaton is massless,
during the period of oscillations, the inflaton has an effective mass 
\beq
m^2_{\rm eff} = 4 \lambda \phi^2 = 4 \sqrt{3 \lambda \rho_{\bar \phi}} \, .
\label{effmass}
\eeq
At $a=\aend$, the inflaton mass is\footnote{This is obtained numerically by solving for the evolution of $\phi$ in the Starobinsky potential and is slightly lower than what is obtained from Eq.~(\ref{effmass}) using $\rho_{\bar \phi} = \rho_{\rm end}$} ${\hat m} = 3.0 \times 10^{13}$~GeV. 
Prior to fragmentation, the effective mass decreases as $m_{\rm eff} \propto a^{-1}$, whereas after fragmentation, it decreases as 
$m_{\rm eff} \propto a^{-1+\frac{b}{4}}$, and we can write
\begin{align}
    m_{\rm eff} & = {\hat m} \left( \frac{\aend}{a} \right) \qquad a < a_\beta \, , \label{effma1} \\
     m_{\rm eff} & = {\hat m} \xi_0^\frac14 \left( \frac{\aend}{a_\beta} \right) \left( \frac{a_\beta}{a} \right)^{1+\frac{b}{4}} \qquad a > a_\beta \, .
     \label{effma2}
\end{align}
The small discontinuity at $a_\beta$ is a result of the approximation of a sudden onset of fragmentation and setting $\rho_{\bar \phi} + \rho_{\delta \phi} \approx \rho_{\delta \phi}$.
The varying effective mass for $k=4$ is distinct from the case for $k=2$, where the inflaton mass is constant.
The fact that the inflaton mass decreases will have a profound effect in the generation of the baryon asymmetry as decays of the inflaton to RHNs will become kinematically forbidden at some point after inflation ends. Specifically, the kinematic shutoff of $\phi \rightarrow NN$ occurs at $a = a_*$, determined by
\beq
{\hat m} \left(\frac{\aend}{a_*} \right) = 2 M_N \, .
\label{Eq:aStar}
\eeq
Then $a_* < a_\beta$ so long as $M_N > 1.6 \times 10^{11}$~GeV.  
 In this case, the evolution history of $\rho_N$ and $n_{B-L}$ is independent of the fragmentation process. We will consider $M_N = 10^{12}$~GeV throughout this work, which corresponds to
 $a_* \simeq 15 \aend$. 

If we neglect the effects of fragmentation,
inflaton decay to SM particles generates a thermal bath with temperature $\Tmax$ as discussed above. The energy density of the radiation then redshifts as $\rho_{\rm R} \propto a^{-\frac{6(k-1)}{k+2}}$ corresponding to $a^{-3}$ for $k=4$. The resulting reheating temperature when the effects of fragmentation are ignored is given by \footnote{Note that we have not included the effects of kinematic suppression which will reduce the value of the effective coupling by about a factor of 2 for $k=4$.  For more detail, see \cite{GKMO2}. }
\beq
\alpha_{\rm RH} \Trh^{4~\rm no~frag} \simeq \frac{1}{3 \pi^4} \lambda y_{\phi ff}^8 M_P^4 \simeq \frac{{{\hat m}}^4 M_P^4}{144 \pi^4  \rho_{\rm end}} y_{\phi ff}^2 \, ,
\label{trh4nofrag0}
\eeq
or
\beq
\Trh^{\rm no~frag} \simeq 4.2 \times 10^{14}~{\rm GeV} y_{\phi ff}^2 \, ,
\label{trh4nofrag}
\eeq
as discussed in Appendix \ref{sec:frag}.

The persistence of a component of the inflaton density in the form of a condensate allows for the inflaton (particle and condensate) to decay as its effective mass remains non-zero. For $a> a_\beta$, however, $\rho_{\rm R} \propto a^{-(3+\frac{b}{2})}$. Reheating is then determined when $\rho_{\rm R} = \rho_{\delta \phi}$, and it is not hard to show that the reheating temperature will scale as $y_{\phi ff}^{\frac{4}{2-b}}$. 
Here, we will make use of the numerical results found in \cite{Garcia:2023dyf}, providing a relation between $y_{\phi ff}$ and $\Trh$. 
For example, for $k=4$,
we take 
\beq
\Trh \simeq 2.6 \times 10^{17}~{\rm GeV} y_{\phi ff}^{5.7} 
\, ,
\label{trhr4frag}
\eeq which includes the effects of fragmentation.
For more detail and analytic approximations on the evolution of the radiation density after fragmentation see Appendix \ref{sec:frag}.

The effects of fragmentation are minor when the inflaton predominantly decays to scalar final states. For decays to scalars, the reheating temperature is \cite{GKMO2}
\begin{align}
   \alpha_{\rm RH} \Trh^4 & = \frac{3}{400 \lambda \pi^2} \mu^4 \qquad \qquad k=2
   \label{trhs2}\, ,\\
    \alpha_{\rm RH} \Trh^4 & = \frac{1}{2^\frac83 3^\frac73 \lambda^\frac13 \pi^{\frac43}} \mu^\frac83 M_P^\frac43 \qquad \qquad k=4 \, ,
      \label{trhs4}
\end{align}
where $\mu$ is the dimensionful coupling of the inflaton to two real scalars. 

We note that the effects of fragmentation can also be avoided if there is in addition to the potential given by Eq.~(\ref{Starok}), a bare mass term for the inflaton \cite{Clery:2024dlk}.  If this is sufficiently large, harmonic oscillations may take over before fragmentation can begin. In this case fragmentation is suppressed and reheating can proceed. However, in that case, we return effectively to the calculation resulting in Eq.~(\ref{dilute}) and a large suppression of the baryon asymmetry. 

In what follows, we concentrate our attention to the cases with $k=2$ and $k=4$ and reheating via inflaton decays to fermions. Later we will comment on the effect of higher values of $k$ and inflaton decays to scalars.

\section{The Boltzmann equations}\label{sec:fermion}

To determine the baryon asymmetry, we need to track several quantities. These include the energy density of the inflaton $\rho_\phi$, the energy density of radiation $\rho_{\rm R}$ produced by both inflaton decays and right-handed neutrino (RHN) decays, and the energy density of RHNs produced non-thermally from inflaton decays as well as from the thermal bath. RHN decays will also lead to a lepton asymmetry and hence a $B-L$ asymmetry which will be related to the baryon asymmetry \cite{spha2,HT} 
so long as sphaleron transitions remain in equilibrium.
In our treatment, reheating is ultimately dominated by inflaton decays to fermions $\phi\rightarrow ff$, though there is some contribution from RHNs decaying into charged leptons and the Higgs boson.

We consider the inflaton potential given by Eq.~(\ref{Starok}) which is approximated by the potential (\ref{appk}) near the minimum at $\phi = 0$. 
This leads to an equation of state parameter $w_\phi = (k-2)/(k+2)$ during inflaton oscillations.
The evolution of the various energy densities of interest is determined by the following set of Boltzmann equations \cite{Giudice:2003jh,Zhang:2023oyo} \begin{equation}
 \begin{aligned}
   &aH \rho_\phi ^\prime +3(1+w_\phi)H\rho _\phi =-(\Gamma_{\phi\rightarrow ff}+\Gamma_{\phi\rightarrow NN}) (1+w_\phi)\rho_\phi \, ,
   \\& aH \rho_{\rm R}^\prime +4H\rho _{\rm R} =\Gamma_{\phi\rightarrow ff}(1+w_\phi)\rho_\phi+\left<\Gamma_N\right> (\rho_N -\rho_N^{\rm eq}) \, , \\
   &  aHn^\prime _N+3H n_N= (1+w_\phi) \Gamma_{\phi\rightarrow NN}\frac{\rho_\phi}{m_\phi}-\left<\Gamma_N\right>\left(n_N-n_N^{\rm eq}\right) \, , \\
      & a H n_{B-L}^\prime +3H n_{B-L}=-\left<\Gamma_N\right>\left[ \epsilon  (n_N -n_N^{\rm eq}) \right]  - \frac12 \left<\Gamma_{\rm ID}\right> n_{B-L} \, ,\\
   &3M_P^2 H^2=\rho_\phi+\rho_{\rm R}+\rho_N \, ,
   \label{Eq:Boltz}
 \end{aligned}
\end{equation}
where $\prime$ denotes a derivative with respect to the scale factor, $a$.  Note that we have elected to evolve the RHN number density $n_N$ rather than the corresponding energy density $\rho_N$. The reason for this is because the equation of state of the RHNs will generally not be a constant throughout the reheating process. However, we will work in controlled regimes where $\rho_N$ can be reconstructed from $n_N$ when necessary. For $k=2$, a close approximation for the RHN energy density is $\rho_N=\langle E_N \rangle m_N$ with $\langle E_N \rangle =0.6 (m_\phi/2)$. We derive this estimate in Appendix \ref{aveE2} and discuss this further in Section \ref{k2}. For $k=4$ where kinematic shutoff happens at $a_*/a_{\rm end} \simeq \mathcal{O}(10)$, taking $\rho_N=M_N n_N$ will be very accurate for $a \gg a_*$.
On the left hand side of the first and second equations, the coefficient of the energy densities is $3 H (1+w)$. On the right hand side of the first three equations, the coefficient of each rate carries a factor of $(1+w)$ as well. The inflaton equation carries only sinks as inflatons are lost through decays and we ignore any inflaton production or losses through scattering. It will not matter whether the inflaton is in the form of a condensate or quanta for $k=4$. The second equation contains sources for radiation through inflaton and RHN decays, where $\left<\Gamma_N\right>$ is the average RHN decay rate,

\begin{equation}
    \left<\Gamma_N\right>=\frac{y_N^2}{8\pi } \frac{M_N^2}{\langle E_N \rangle} \qquad k=2 \, ,\label{RHNgamma0}
\end{equation}
\begin{equation}
    \left<\Gamma_N\right>=\frac{K_1(M_N/T)}{K_2(M_N/T)}\frac{M_N}{8\pi }y_N^2 \qquad k=4  \, ,\label{RHNgamma}
\end{equation}
where the factor of $M_N/\langle E_N \rangle$ accounts for the time dilation of $N$ decay (see Appendix \ref{aveE2}) and 
where $K_i$ is the Bessel function of $i$-th kind\footnote{This rate assumes Maxwell-Boltzmann statistics and is valid for $M \gtrsim T$.}. The second equation in (\ref{Eq:Boltz}) also contains a sink for the equilibrium production of RHNs. 
The sources and sinks for the third equation are determined by the previous two. In the fourth equation of (\ref{Eq:Boltz}), the $B-L$ asymmetry is related to the density of RHNs through the parameter $\epsilon$ which characterizes the net $L$ asymmetry produced by a RHN decay (thus enters with a minus sign for $n_{B-L}$). The last term in the fourth equation corresponds to removing the asymmetry through inverse decays and $\Gamma_{\rm ID} = \Gamma_N n_N^{eq}/n_l^{eq}$. The final equation in (\ref{Eq:Boltz}) is simply the Friedmann equation for which all energy densities contribute to the expansion rate, $H$, and the inflaton will be the dominant contributor throughout the reheating process.

For $k=2$, the inflaton mass and decay rate are fixed. For larger  $k$, the decay rate is non-constant and determined by the effective mass of the inflaton which for $k=4$ 
is given in Eq.~(\ref{effmass}).  The decay rate in terms of the effective mass is used in the Boltzmann equations.
For $k=2$, we can write 
\begin{equation}
    \Gamma_{\phi\rightarrow NN}= \frac{y_{\phi NN}^2 m_\phi }{8\pi} \sqrt{1-\frac{4 M_N^2}{m_\phi^2}} = \frac{y_{\phi NN}^2\sqrt{\lambda}M_P}{8\pi} \sqrt{1-\frac{4 M_N^2}{\lambda M_P^2}} \, , 
    \label{k2boltzN}
\end{equation}
where $y_{\phi NN}$ is the inflaton coupling to $NN$. 
For $k=4$, as discussed above, 
if $M_N \gtrsim 10^{11}$~GeV, inflaton decays to $N$ occur before fragmentation, and we can write
\begin{equation}
    \Gamma_{\phi\rightarrow NN}(a)= \frac{y_{\phi NN}^2 m_{\rm eff}(a) }{8\pi} \sqrt{1-\frac{4 M_N^2}{m_{\rm eff}(a)^2}} \simeq \frac{y_{\phi NN}^2(3\lambda\rho_{\phi}(a))^{1/4}}{4\pi} \sqrt{1-\frac{M_N^2}{\sqrt{3 \lambda \rho_{\phi}(a)}}} \, .
    \label{k4boltzN}
\end{equation}
We have explicitly included the scale factor dependence for the relevant quantities, since the evolution of $\Gamma_{\phi \rightarrow NN}(a)$ is of fundamental importance for our mechanism of interest. Similar expressions for inflaton decays to fermions can be written in terms of the coupling $y_{\phi ff}$.

\section{The evolution of the baryon asymmetry for $k=2$}
\label{k2}

The problem for leptogenesis in low reheating scenarios is particularly acute for matter-like reheating ($k=2$, $w_\phi=0$). 
In this section, we will assume the standard Starobinsky potential in Eq.~(\ref{Staro}) and
show how the baryon asymmetry is suppressed as $\Trh$ is decreased, ultimately with 
a non-recoverable dilution factor given by Eq.~(\ref{dilute}), with $\gamma = 5$ for $k=2$.

In the introduction, we discussed the basic features of non-thermal leptogenesis in the case where one assumes that the inflaton decays instantaneously at the end of inflation with non-zero branching fractions to both RHNs and SM radiation. The RHNs subsequently decay into SM particles, leading to a lepton asymmetry which is converted into a baryon asymmetry via sphaleron transitions. This led us to a rough estimate of the final baryon asymmetry in the instantaneous reheating approximation, given by Eq.~(\ref{Eq:instRehApprox}). In this section and throughout the remainder of this work, we will treat the reheating period more carefully as a non-instantaneous process.

A key aspect of leptogenesis for $k=2$ is that nonthermal production of RHNs via inflaton decays will continuously occur throughout the entirety of reheating, which will not be the case for $k \geq 4$. The reason for this is that the decay rate $\Gamma_{\phi \rightarrow NN}$ given in Eq.~(\ref{k2boltzN}) is simply proportional to $m_\phi$, which is constant for $k=2$ ($m_\phi=3.0\times10^{13}$~GeV). Therefore, even for low reheating temperature scenarios where the energy density of the inflaton condensate drops quite low (for instance $\rho_{\phi}$ approaches $(1 \text{ GeV})^4$ for $T_{\rm RH}=1$~GeV), the inflaton is still able to produce high energy decay products with masses much greater than the reheating temperature. Specifically, RHNs with mass $M_N=10^{12}$~GeV can still be efficiently produced late in reheating. For $k\geq 4$, the inflaton's effective mass will instead quickly drop below the kinematic threshold required for RHN production, which will qualitatively change the dynamics. For $k=2$, the continuous production of RHNs throughout reheating will interestingly prove to be a barrier to successful leptogenesis, rather than an asset.

Before proceeding with our numeric results for $k=2$, we will first refine our analytic estimate of the final baryon asymmetry. We first notice that the simple estimate in Eq.~(\ref{Eq:instRehApprox}) relies upon the heuristic relation, $Br = \frac{\Gamma_{\phi NN}}{\Gamma_\phi,\rm tot} \sim \frac{n_N}{n_R}$. While this relation is valid for $\Gamma_{\phi \rightarrow ff}\gg \Gamma_{\phi\rightarrow NN}$ the instant after the inflaton decays, the process of thermalization (which we assume to be instantaneous throughout this work) will immediately change the radiation number density such that the above heuristic is no longer accurate even in the instantaneous reheating approximation. In particular, the energy of inflaton decay products for $1\rightarrow 2$ decays for both RHNs and SM radiation will be $E=m_\phi/2$ immediately after decay, such that the moment after inflaton decay (prior to thermalization) we will have $\rho_{\rm R}\propto(m_\phi/2) n_{R,i}$ and $\rho_N=(m_\phi/2) n_{N}$. After thermalization, however, we will have $\rho_{\rm R}=\alpha T^4\propto Tn_{R,f}$, where the initial and final radiation number densities differ by a factor of $T/(m_\phi/2)$. Note that for the parameter choices we use in this study, the RHNs will not be thermalized during the relevant epochs, so the RHN energy density remains $\rho_N \simeq(m_\phi/2) n_{N}$. Therefore, with thermalization taken into account, the ratio of the energy densities at the completion of reheating is related to the branching ratio and we can determine a relation between $n_N$, $T_{\rm RH}$, and $Br$ as follows: 
\beq
Br
\sim
\frac{\Gamma_{\phi NN}}{\Gamma_{\phi ff}}
\sim
\left.\frac{\rho_N}{\rho_{\rm R}}\right|_{a=a_{\rm RH}}=\left.\frac{\langle E_N \rangle n_N}{\rho_{\rm R}} \right|_{a=a_{\rm RH}}
\approx
\frac{(m_\phi/2)n_N(a_{\rm RH})}{\alpha T_{\rm RH}^4} \,,
\eeq where we have taken the average RHN energy at reheating to be $\langle E_N \rangle\approx(m_\phi/2)$ (we derive a more precise relation for $\langle E_N \rangle$ in Appendix \ref{aveE2}). 
 Thus, we can refine the crude estimate of the baryon asymmetry in Eq.~(\ref{Eq:instRehApprox}) to the following
\beq
Y_{B,\infty} \equiv \frac{n_B}{s}|_{a=a_\infty} \sim \frac{8}{23} \epsilon \frac{n_N(a_{\rm RH})}{s(a_{\rm RH})}\simeq \frac{8}{23} \frac{3}{4} \epsilon Br  \left(\frac{T_{\rm RH}}{m_\phi /2}\right), \hspace{5mm} \text{for } T_{\rm RH} > T_{\rm sph}.\label{Eq:Yb}
\eeq 
Note that while the result in Eq.~(\ref{Eq:Yb}) does not explicitly depend upon the RHN decay rate (which in turn depends on $y_N$ and $M_N$), this result does require the basic restrictions that the RHNs decay prior to sphaleron decoupling and that RHNs do not transiently dominate the energy density of the universe. 

In the above discussion, we assumed that RHNs decay near the time of reheating or shortly thereafter. We now turn our attention to the case where $y_N$ is sufficiently large such that RHNs begin decaying well before the completion of reheating. Interestingly, we will find that the final asymmetry has the same form as Eq.~(\ref{Eq:Yb}) if $T_{\rm RH}>T_{\rm sph}$.
We will define the temperature $T_N$ and the scale factor $a_N$ to be the values of $T$ and $a$ when $\Gamma_N \sim H$. As long as $\Trh>T_N$, reheating can be regarded instantaneous as it terminates before the RHNs decay. On the other hand, when $\Trh<T_N$, the RHNs start to decay during reheating while still being continuously generated by the inflaton. The evolution of the baryon asymmetry in this scenario will differ from the instantaneous case. From the Boltzmann equations \eqref{Eq:Boltz} and using $H=H_{\rm end}(\aend/a)^{3/2}$ we find that $n_N\propto a^{-3/2}$ for $\aend \lesssim a \lesssim a_{N}$, and when $\Gamma_N \approx H$, the RHNs begin to decay and $n_N\propto a^{-3}$ for $a_{N} \lesssim a \lesssim a_{\rm RH}$.\footnote{Note that $n_N$ does not drop off exponentially as RHNs are continuously being produced by inflaton decays.} 
The 4th Boltzmann equation in \eqref{Eq:Boltz} is then easily solved to give
$n_{B-L}\propto a^{-3/2}$ for $\Trh<T<T_N$. Since the temperature  evolves as $a^{-3/8}$, we find that the baryon asymmetry is diluted as $Y_B\propto T/T_N$ until  $\Trh$ if $\Trh>T_{\rm sph}$ or until  $T_{\rm sph}$ if $\Trh<T_{\rm sph}$.

This dilution phase exists for $\Trh<T_N$, namely when 
\begin{equation}
    \Trh \lesssim \left(\frac{90 M_P^2 {\Gamma_N^2}}{g_{\rm RH}  {\pi^2  }}\right)^{\frac{1}{4}} \, .
\end{equation}
For example, using Eq.~(\ref{RHNgamma0}) for $\Gamma_N$ and Eq.~(\ref{aveEN}) for $\left< E_N \right>$ with $y_N=10^{-4}$ and $M_N=10^{12}$ GeV, we get $\Trh\lesssim 4.6\times 10^9$ GeV. 
Then for $\Trh < T_N$, we expect
\beq
 Y_B|_{T=T_{\rm sph}}  \simeq  \frac{8}{23} \frac{3}{4} \epsilon Br \left(\frac{T_{N}}{m_\phi /2}\right) \left(\frac{T_{\rm RH}}{T_N}\right) \qquad T_N > \Trh > T_{\rm sph} \,\label{r3},
\eeq such that the final asymmetry in this case is equivalent to Eq.~(\ref{Eq:Yb}).
Finally for $\Trh < T_{\rm sph}$, we must include the dilution from Eq.~(\ref{dilute}) so that
\beq
 Y_B|_{T=T_{\rm RH}}  \simeq  \frac{8}{23} \frac{3}{4} \epsilon Br \left(\frac{T_N}{m_\phi /2}\right) \left(\frac{T_{\rm sph}}{T_N}\right) \left(\frac{T_{\rm RH}}{T_{\rm sph}}\right)^5 \qquad T_{\rm sph} > \Trh \, . \label{r40} 
\eeq which simplifies to
\beq
 Y_B|_{T=T_{\rm RH}}  \simeq  \frac{8}{23} \frac{3}{4} \epsilon Br \left(\frac{T_{\rm sph}}{m_\phi /2}\right)  \left(\frac{T_{\rm RH}}{T_{\rm sph}}\right)^5 \qquad T_{\rm sph} > \Trh \, . \label{r4}
\eeq
In summary, the decay temperature of the RHNs, $T_{N}$, never dictates the final asymmetry as long as $T_N > T_{\rm sph}$, despite the fact that it will change the precise shape of the $n_{B-L}$ history.

\begin{figure}[h!] \includegraphics[width=\textwidth]{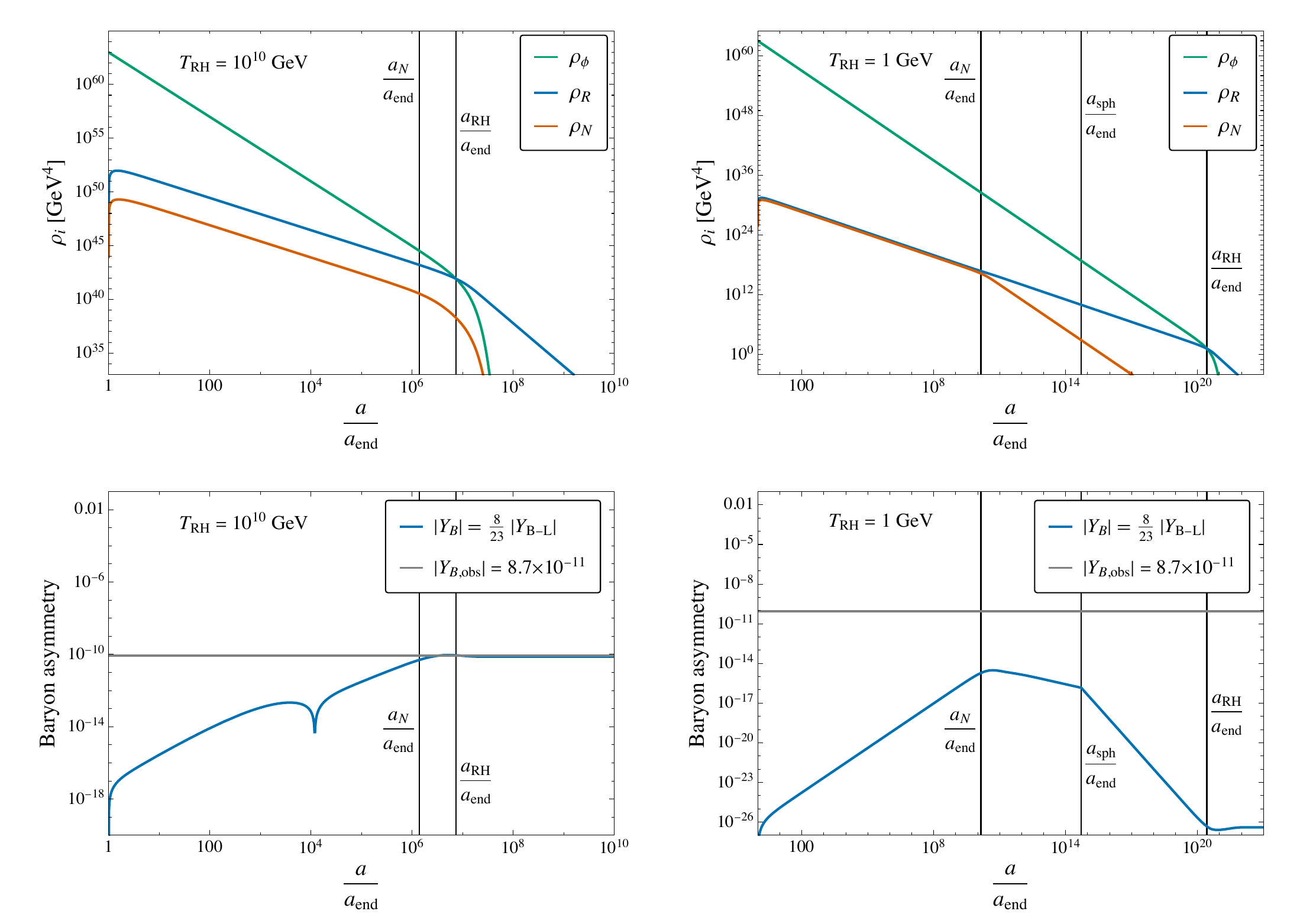} 
    \caption{The evolution of energy density components (top panels) and the baryon asymmetry (bottom panels) for $k=2$ and $\Gamma_{\phi \rightarrow NN}\leq\Gamma_{\phi \rightarrow ff}$. The left panels correspond to $T_{\rm RH}=10^{10}$~GeV while the right panels depict $T_{\rm RH}=1$~GeV. In both cases, $M_N=10^{12}$~GeV and $|\epsilon|=10^{-4}$. Other parameters for the left panels are: $y_N=10^{-3}$, $y_{\phi ff}=2.1\times10^{-5}$, and $y_{\phi NN}=1.5\times 10^{-6}$. For the right panels, the remaining parameters are $y_N=10^{-6}$ and $y_{\phi ff}=y_{\phi NN}=1.2 \times 10^{-15}$. The horizontal gray lines in the bottom panels depict the observed value of the baryon asymmetry. The vertical black lines correspond to values of the scale factor at $N$ decay, sphaleron decoupling and reheating.}
    \label{k2plotv1}
\end{figure}

We are now in a position to present our numerical results for $k=2$. Let us first illustrate the evolution of the energy density components and the baryon asymmetry in the setting of high reheating temperature. We initially restrict ourselves to the case where $\Gamma_{\phi \rightarrow ff}\geq\Gamma_{\phi \rightarrow NN}$, and we will consider the alternative below. 
As discussed in the previous section, the momentum distribution of the RHNs will typically be non-thermal for most of the evolution for $k=2$, since RHNs with energy $E_N \sim m_\phi/2$ are continuously produced by inflaton decays long after the temperature drops below $T = M_N$. 
We use Eq.~(\ref{RHNgamma0}) for the decay rate and Eq.~(\ref{aveEN}) for the average energy per particle. 
For relatively high reheating temperatures satisfying the hierarchy $ T_{\rm RH}\gg T_{\rm sph}$, the estimate of the final baryon asymmetry is well-approximated by Eq.~(\ref{Eq:Yb}) (or equivalently (\ref{r3})), and it is not difficult to find values of $Br$ and $\epsilon$ for which the observed asymmetry is satisfied. This scenario is depicted in the left panels of Fig.~\ref{k2plotv1}, where the choices of $y_{\phi NN}=1.5\times 10^{-6}$ and $y_{\phi ff}=2.1\times10^{-5}$ (corresponding to $Br \simeq 0.005$), and $|\epsilon| = 10^{-4}$ are expected to approximately yield the observed asymmetry
from Eq.~(\ref{Eq:Yb}). 
The evolution of radiation and RHNs are parallel and fall off as $a^{-3/2}$ during reheating until the RHNs decay (shown here to be slightly before the inflaton efficiently decays at reheating). In the bottom left panel of Fig.~\ref{k2plotv1}, we depict the evolution of the baryon asymmetry compared to the observed asymmetry. The complicated history is a result of the competition between RHN decays and inverse decays (see the next section for more detail), though the final asymmetry is simply determined by the total number of RHNs that decay prior to sphaleron decoupling. 

While the observed baryon asymmetry is easy to obtain for high reheating temperatures, one quickly encounters difficulties if one attempts to do the same for low reheating temperatures with $k=2$. In the right panels of Fig.~\ref{k2plotv1}, we depict a similar scenario described above but for $T_{\rm RH}=1$~GeV. In this case, we encounter two challenges which directly correspond to the two suppression factors in Eq.~(\ref{r4}). The first difficulty is the suppression of the asymmetry by a factor of $T_{\rm sph}/(m_\phi/2) \sim 10^{-11}$. Since we are assuming $Br < 1$, and thus $(8/23)(3/4) \epsilon Br < 1$, there is no possibility to obtain the observed asymmetry, when using the standard type 1 see-saw, which typically requires $|\epsilon| \lesssim 10^{-3}$. Indeed for the parameters chosen for the right panels of Fig.~\ref{k2plotv1},
$Br = 1/2$, and even without the suppression for 
$T < T_{\rm sph}$, the baryon asymmetry is $\mathcal{O}(10^{-16})$ for $\Trh = T_{\rm sph}$. The second challenge is that for $\Trh = 1$~GeV, there is an additional suppression of $(\Trh/T_{\rm sph})^5 \sim 10^{-11}$, leaving the final asymmetry at $Y_B < 10^{-26}$ as seen in the lower right panel of the figure.  As a result, obtaining the correct baryon asymmetry for low reheating temperatures with $k=2$ and $\Gamma_{\phi \rightarrow ff}\geq\Gamma_{\phi \rightarrow NN}$ is not possible for ordinary leptogenesis.

Perhaps the next approach to try to overcome the challenges described above is to increase $\Gamma_{\phi \rightarrow NN}$ relative to $\Gamma_{\phi ff}$, allowing for the possibility that $\rho_N > \rho_{\rm R}$, with the aim of increasing the lepton asymmetry which can be converted to a baryon asymmetry at $T_{\rm sph}$. In particular, one can allow the RHN density to exceed the radiation density during the reheating period, which is distinct from the scenario explored in Fig.~\ref{k2plotv1}.~\footnote{This type of evolution was considered in the context of freeze-in in \cite{Cosme:2024ndc}.} We illustrate this alternative hierarchy ($\rho_{\rm N,max}>\rho_{\rm R,max}$) for both high ($T_{\rm RH}=10^{10}$~GeV) and low ($T_{\rm RH}=1$~GeV) reheating temperatures in Fig.~\ref{k2plot} below. For these cases, we take $y_{\phi ff}=0$, such that the SM radiation is no longer produced by inflaton decays but rather solely via RHN decays \footnote{If we were to take $y_{\phi ff}$ to be non-zero but still less than $y_{\phi NN}$, reheating would simply occur earlier and at higher temperatures for the same value of $y_{\phi NN}$, which is counterproductive to our purposes in exploring this alternative hierarchy for low reheating temperatures.}.
Note in this case, $Br \approx 1$, and the analytic approximation in Eq.~(\ref{Eq:Yb}) is no longer accurate. We depict the numerical evolution of the energy density components and the baryon asymmetry for two sets of parameter choices. For high reheating temperatures (left panels of Fig.~\ref{k2plot}), as we saw before, it is not difficult to obtain the correct baryon asymmetry, while for low reheating temperatures we again encounter difficulties. The first difficulty is again the dilution factor, which sharply penalizes the final asymmetry for reheating temperatures below $T_{\rm sph}$, and this is not altered by increasing the decay rate to RHNs. This dilution penalty on the baryon asymmetry for low reheating temperatures can be seen in the bottom right panel of Fig.~\ref{k2plot} as the sharp decline between $a_{\rm sph}/a_{\rm end}$ and $a_{\rm RH}/a_{\rm end}$. Secondly, we must not allow the RHN density to overtake the inflaton density. If it does, reheating will be due to RHN decay rather than inflaton decay. In this case, a low reheating temperature, $\Trh < T_{\rm sph}$ would require that the majority of RHNs decay after sphaleron decoupling and produce a lepton asymmetry which is not processed into a baryon asymmetry. We therefore restrict ourselves to the case where $\rho_N(a) < \rho_\phi(a)$ for all $a<a_{\rm RH}$. However, this restriction gives us a fundamental upper limit on the maximum density of RHNs, which precludes an asymmetry large enough to overcome the subsequent dilutions. This limit on the maximum density of RHNs is ultimately due to the fact that RHNs are continuously produced throughout reheating for $k=2$. As a result, we find that allowing for $\Gamma_{\phi \rightarrow NN}>\Gamma_{\phi \rightarrow ff}$ does not enable us to circumvent the challenges we first encountered in the $\Gamma_{\phi \rightarrow NN}\leq \Gamma_{\phi \rightarrow ff}$ case. Indeed, we see in the right panels of Fig.~\ref{k2plot} that for a representative parameter set for this alternative hierarchy ($\Gamma_{\phi \rightarrow NN}\gg \Gamma_{\phi \rightarrow ff}$), the final baryon asymmetry is still many orders of magnitude below the observed value.

To summarize, in this section we found that the final baryon asymmetry in our setup is sensitive to two main factors: 1) the number of RHNs which have decayed prior to $T_{\rm sph}$, and 2) any dilution of the baryon asymmetry between the sphaleron transition and reheating. These factors make successful leptogenesis challenging at low reheating temperatures. Because we have found that generating the correct baryon asymmetry is not possible for the simplest realizations of non-thermal leptogenesis with $k=2$, we now proceed to investigate the analogous setup for inflaton potentials with a quartic minimum, namely $k=4$.

\begin{figure}[ht!] \includegraphics[width=\textwidth]{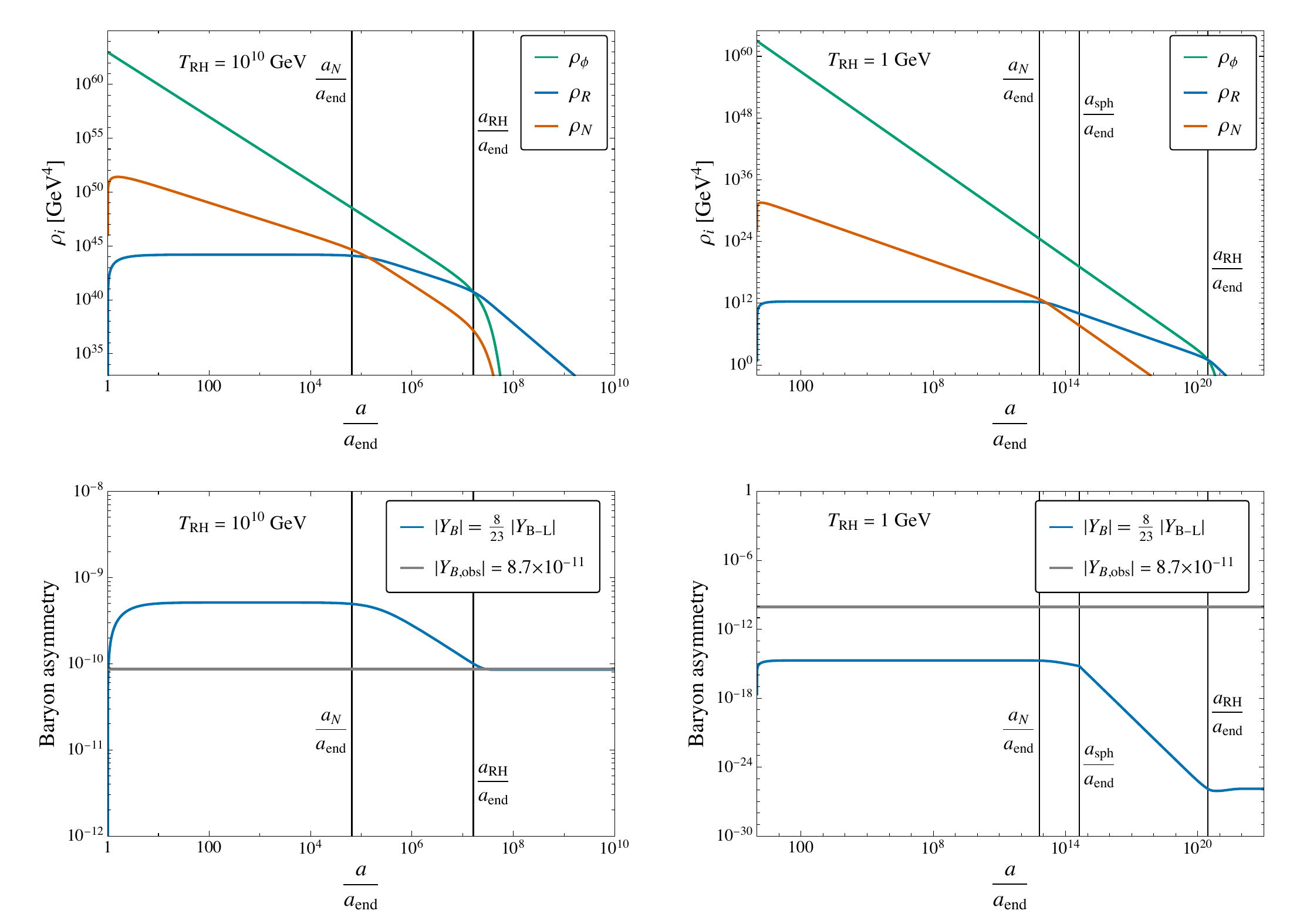} 
    \caption{The evolution of energy density components (top panels) and the baryon asymmetry (bottom panels) for $k=2$ and $\Gamma_{\phi \rightarrow NN}\gg\Gamma_{\phi \rightarrow ff}$. The left-hand panels correspond to a relatively high reheating temperature, $T_{\rm RH}=10^{10}$~GeV, while the right-hand panels illustrate the evolution for $T_{\rm RH}=1$~GeV. The parameter values for the left panels are $y_{\phi NN}=1.8\times 10^{-5}$, $y_{N}=10^{-2}$, and $|\epsilon|=3.5\times 10^{-7}$. The parameter values for the right panels are $y_{\phi NN}=2.1\times10^{-15}$, $y_N=10^{-8}$, and $|\epsilon|=10^{-4}$. In all cases, $M_N=10^{12}$~GeV and $y_{\phi ff}=0$.}
    \label{k2plot}
\end{figure}

\section{The evolution of the baryon asymmetry for $k=4$}
\label{k4}

The salient difference between the $k=2$ case and $k>2$ is the fact that the inflaton's effective mass is no longer constant. As we will see, this has dramatic consequences for non-thermal leptogenesis via inflaton decays to RHNs. In particular, since the inflaton mass decreases with increasing scale factor, the inflaton mass will quickly drop below the RHN mass early in the reheating process, which leads to a kinematic shutoff of the $\phi \rightarrow NN$ channel at a scale factor $a=a_*\simeq \mathcal{O}(10)$. Thus, the final baryon asymmetry for $k=4$ will ultimately trace its origin to an early RHN population which is largely fixed at $a=a_*$, very early in the reheating process. Recall that for $k=2$, RHN production was continuous throughout the entirety of reheating, which led to many additional consequences such as sensitivity to the hierarchy of $\Gamma_{\phi NN}$ vs. $\Gamma_{\phi ff}$ and a limit on the maximum allowable RHN density to preclude $\rho_N$ from overtaking $\rho_\phi$. We will see that for $k>2$, the situation is comparatively much simpler and far more amenable to successful baryogenesis.

For $a<a_*$, the decay rate $\Gamma_{\phi\rightarrow NN}$ for $k=4$ scales as $a^{-1}$ from Eqs.~(\ref{k4boltzN}) and (\ref{effma1}). The decay rate to Standard Model fermions throughout the reheating process, however, is more complicated due to the effects of fragmentation of the inflaton condensate. Prior to fragmentation, the inflaton decay rate to SM fermions is given by an expression similar to that for decays to $N$,
\begin{equation}
     \Gamma_{\phi\rightarrow ff}=\frac{y_{\phi ff}^2m_{\rm eff}}{8\pi} \qquad a < a_\beta \, ,
    \label{k4boltzf}
\end{equation}
with $m_{\rm eff}$ given by Eq.~(\ref{effma1}).
After fragmentation begins, the inflaton energy density is dominated by the free (relativistic) quanta. The decay rate for these are suppressed by a time dilation factor so that
\begin{equation}
     \Gamma_{\delta\phi\rightarrow ff}=\frac{y_{\phi ff}^2m_{\rm eff}^2}{8\pi {\bar E}} \qquad a > a_\beta \, ,
    \label{k4boltzdf}
\end{equation}
with $m_{\rm eff}$ given by Eq.~(\ref{effma2}) and where ${\bar E}$ is the average energy per particle (for more detail, see Appendix \ref{sec:frag}). 
Prior to fragmentation, $\Gamma_{\phi\rightarrow ff}\propto a^{-1}$, whereas subsequently, since ${\bar E}$ scales as $a^{-1}$, $\Gamma_{\phi\rightarrow ff}\propto a^{-(1+\frac{b}{2})}$.  

The maximal temperature during reheating for $k=4$ is given by Eq.~(\ref{tmax4}) and expressed in terms of $y_{\phi ff}$ is simply\footnote{To obtain the numerical value we replaced $\lambda^\frac14$ with 
${\hat m}/2\sqrt{3}\rho_{\rm end}^\frac14$.}
\begin{equation}
    T_{\rm max}
    \simeq  7.2 \times 10^{11}\text{ GeV} \left(\frac{y_{\phi ff}}{1.0 \times 10^{-6}}\right)^{1/2} \, .
    \label{tmaxeq}
\end{equation}
Sphaleron processes will be in equilibrium at  $T_{\rm sph} = 130~{\rm GeV} < T <10^{12}\text{ GeV}$. 
The reheating temperature for $k=4$ is given in Eq.~(\ref{trhr4frag}) when the effects of fragmentation are included. 
In what follows, we will be primarily interested in $4\text{ MeV}<T_{\rm RH}< T_{\rm sph}$ corresponding to $3\times 10^{-4}<y_{\phi ff} < .002$. In this case, sphaleron processes cease before the end of reheating, and from \eqref{tmaxeq} we get $1.3\times 10^{13}\text{ GeV}<T_{\rm max}<3.3\times 10^{13}\text{ GeV}$.

For $M_N \sim 10^{12}$~GeV,  the inflaton mass at the end of inflation is ${\hat m}=3.0\times 10^{13}\text{ GeV}>2M_N$ and the RHN mass is then smaller than $T_{\rm max}$ in the parameter space we consider. More precisely, after inflation, the temperature quickly rises and  reaches $T_{\rm max}$ at $a_{\rm max}=4/3$, and then decreases as $T\propto a^{-3/4}$ until the fragmentation effect becomes important. Therefore, thermally produced RHNs are relativistic for a short  period after inflation (RHNs produced by inflaton decay are also initially relativistic).

When $T\ll M_N$ (shortly after inflation ends), $n_N^{\text{eq}}$ in Eq.~(\ref{Eq:Boltz}) is exponentially suppressed by $e^{-M_N/T}$, so thermal production is negligible.
When $T\gg M_N$, we have
\begin{equation}
     \left<\Gamma_N\right>\simeq \frac{y_N^2M_N^2}{8\pi {\bar E_N}},\quad n_N^{\rm eq} =   \frac{3 \zeta(3) T^3}{2 \pi^2} \, ,
\end{equation}
where ${\bar E_N} \simeq 3.15 T$.\footnote{Maintaining Maxwell-Boltzmann statistics would have given $ \left<\Gamma_N\right>\simeq M_N^2y_N^2/16\pi T $ and $n_N^{\rm eq} \simeq  2 T^3/ \pi^2.$} 
In the high temperature regime, the RHS of the third equation in \eqref{Eq:Boltz} contains two production terms (neglecting factors of $\mathcal{O}(1)$ in the thermal rate):
\begin{equation}
\begin{aligned}
    &\text{non-thermal:}\quad \frac{y_{\phi NN}^2\rho_\phi}{ 6 \pi }\simeq 4 \times 10^{61}\text{ GeV}^4 y_{\phi NN}^2 \left(\frac{\aend}{a} \right)^4,\\  &\text{thermal:}\quad \frac{M_N^2y_{N}^2}{8\pi T} \frac{T^3}{\pi^2} \simeq 4\times 10^{47}\text{ GeV}^4y_N^2\left(\frac{T}{10^{13}\text{ GeV}}\frac{M_N}{10^{12}\text{ GeV}}\right)^2,
    \label{Eq:productionTerms}
\end{aligned}
\end{equation}
where we immediately notice the significant difference in the normalization of the two rates. Recall that the non-thermal production of $N$ ceases when the decay $\phi\rightarrow NN$  is kinematically forbidden at $a_*$. Furthermore, the thermal production will quickly become Boltzmann suppressed for $T \lesssim M_N$. As a result, for the parameters we consider in this work, the non-thermal production will always be dominant, and the non-thermal term will typically be larger than the thermal term by many orders of magnitude. In this case, the RHN production is mainly sourced by the inflaton decay.  In the following analytical arguments, we will work in the regime $y_N \simeq y_{\phi NN}$, $M_N\simeq 10^{12}$ GeV, and we neglect the thermal production term and we take the RHN to be approximately non-relativistic. Because all of the relevant RHN production for $k=4$ occurs prior to $a= a_*\sim 15 a_{\rm end}$, while $a_{\rm RH}$ is frequently greater than $\mathcal{O}(10^{12})$ for the low reheating temperatures we consider, non-relativistic RHN is indeed a very good approximation. Our numerical results include both the thermal and non-thermal production channels. 

If we neglect the thermal term for $a<a_*$,
the analytic solution of the third equation in \eqref{Eq:Boltz} is:
\begin{equation}
    n_N(a<a_*)=\frac{\sqrt{3 \rho_{\rm end}} M_P y_{\phi NN}^2}{6\pi } \left( \frac{\aend}{a} \right)^2 \left( 1-\frac{\aend}{a} \right) \, .
    \label{solnnainf}
\end{equation}
We can see that after inflation, $n_N$ quickly increases and reaches its maximum at $1.5 \aend $. Then, $n_N$ redshifts as $a^{-2}$ while it is being produced nonthermally via inflaton decays until these decays become kinematic inaccessible at $a_*$ due to the evolving effective mass of the inflaton. 
For  $a\gg a_*$, we have $M_N\gg T$ so the decay rate becomes:
\begin{equation}
\left<\Gamma_N\right> \simeq \frac{M_N}{8\pi}y_N^2 \, .
\label{gammansmallt}
\end{equation}  
Note that at $a_*$ the temperature is $7\times 10^{12}$ GeV for $y_{\phi ff} = 0.002$, and therefore cannot be neglected compared to $M_N$. Accordingly, for $M_N=10^{12}$ GeV, the decay rate in \eqref{gammansmallt} should be corrected by  a factor $c\sim \mathcal{O}(0.01-0.1)$ near $a_*$. When $a\gg a_*$, we recover \eqref{gammansmallt} with $c=1$. For the purposes of obtaining a reasonably simple analytic result, we assume that  $\left<\Gamma_N\right> $ is constant.
Using the boundary condition at $a_* $ from \eqref{solnnainf}, we get
\begin{equation}
    n_N(a>a_*)=\frac{\sqrt{3 \rho_{\rm end}} M_P y_{\phi NN}^2}{6\pi } \left(\frac{a_*}{a} \right)^3 \left( \frac{\aend}{a_*} \right)^2 \left( 1-\frac{\aend}{a_*} \right) e^{\frac{c M_N y_N^2}{16\pi H_{\rm end}\aend^2}(a_*^2-a^2)} \, .
    \label{solnnasup}
\end{equation}

\begin{figure}[h!] \includegraphics[width=.7\textwidth]{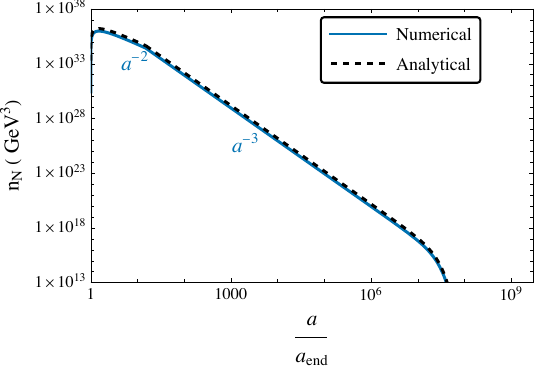} 
    \caption{The evolution of $n_N$ as a function  of $a/\aend$ for $k=4$. We have chosen $y_{\phi ff}=0.002$, $y_N=y_{\phi NN}=10^{-6}$, $M_N=10^{12}$~GeV. The analytical curve corresponds to the solutions in Eqs.~\eqref{solnnainf} and \eqref{solnnasup}. 
    }
    \label{nnplot}
\end{figure}

For the numerical evolution of the RHN number density, we use \eqref{Eq:Boltz}  with
\beq
H\simeq\frac{\rho_{\rm end}^\frac12  }{\sqrt{3} M_P}  \left(\frac{a_{\rm end}}{a} \right)^2 \, ,
\eeq
which is a good approximation up to $a_\beta$, and the decay rate given in \eqref{RHNgamma}.  In Fig.~\ref{nnplot}, we compare the analytic solutions above with the numerical solution of $n_N$, which are in very good agreement.   As is expected from the analytic arguments, $n_N$ redshifts as $a^{-2}$ from $1.5 \aend$ to $a_* \approx 15 \aend$, then as $a^{-3}$ until the exponent in \eqref{solnnasup} becomes $\mathcal{O}(1)$ (when $\Gamma_N \simeq H)$ and RHN decays become rapid at $a_{N}$. More precisely, we define $a_N$ when the exponent in \eqref{solnnasup} is equal to $-1$ (assuming $c=1$):
\begin{equation}
    \frac{a_N}{\aend}=\sqrt{\frac{16\pi H_{\rm end} }{M_N y_N^2}} \, .
\end{equation}
The parameter choices used in Fig.~\ref{nnplot} lead to $a_N\simeq 2\times 10^7 \aend$, which matches the cutoff point.

Next, if we ignore the thermal terms, $n_{B-L} $ follows from the equation:
\begin{equation}
    a H n_{B-L}^\prime +3H n_{B-L}=-\left<\Gamma_N\right> \epsilon   n_N \, .
    \label{nblequation}  
\end{equation}
The analytic solution to this equation is 
\begin{equation}
    n_{B-L}(a<a_*)=-\frac{\epsilon c M_N M_P^2 y_N^2y_{\phi NN}^2}{16 \pi^2 } \left[ \frac13\left( \frac{a}{\aend}\right)^3 - \frac12 \left( \frac{a}{\aend} \right)^2 + \frac16 \right] \left(\frac{\aend}{a} \right)^3 \, ,\label{nblsmalla}
\end{equation}
and\begin{equation}
    n_{B-L}(a>a_*)=\frac{a_*^3}{a^3}\left[  n_{B-L}(a_*)+\epsilon n_N(a_*)\left(e^{\frac{(a_*^2-a^2) cM_N y_N^2}{16\pi H_{\rm end}\aend^2}}-1\right)\right] \, . \label{nblbiga}
\end{equation}
Soon after $a_*$, we can expand \eqref{nblbiga} to obtain $n_{B-L}\propto a^{-1}$. For $a>a_{N}\gg a_*$, $n_{B-L}\propto a^{-3}$.

\begin{figure}[h!] \includegraphics[width=.7\textwidth]{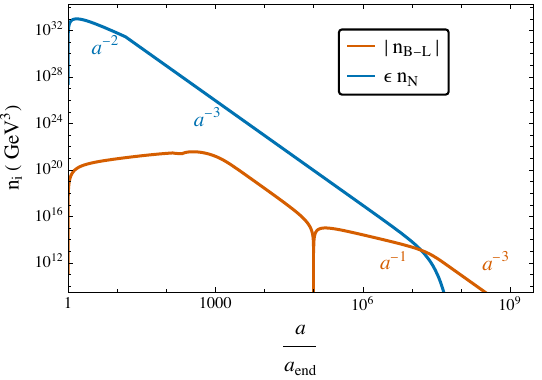} 
    \caption{$|n_{B-L}|$ and $\epsilon n_N$ as functions of $a/\aend$ with the asymptotic behaviors as labeled. The absolute value on $n_{B-L}$ accounts for the change in sign of $B-L$.   We have chosen $\epsilon=-10^{-3}$, $y_{\phi ff}=0.002$, $y_N=y_{\phi NN}=10^{-6}$, $M_N=10^{12}$~GeV.}
    \label{nikk4}
\end{figure}

Note that at high temperatures, we can no longer neglect the equilibrium number density $n_N^{\rm eq}$ in the Boltzmann equation (\ref{Eq:Boltz}). If initially $T> M_N$, and $n_N<n_N^{\rm eq}$, then $n_{B-L}$ will be driven negative, so our analytic formula \eqref{nblsmalla} does not hold. However, as the temperature drops below $M_N$ at later times, $n_N^{\rm eq}$ becomes negligible, leading to positive $n_{B-L}$, and the  qualitative behavior from  \eqref{nblbiga} still holds (though with different boundary conditions), namely $n_{B-L}\propto a^{-3}$ once $n_N$ decays. In Fig.~\ref{nikk4}, we compare the numerical solutions of $n_{B-L}$ and $\epsilon n_N$ in this case. $n_{B-L}$ is initially negative, and when $T\lesssim M_N$, the inverse decay become subdominant, and $n_{B-L}$ increases, becoming eventually positive. The late-time  asymptotic behavior matches the analytic arguments from \eqref{nblbiga}. When the RHNs decay at $a_N$, we also have the relation $n_{B-L}=\epsilon n_N$ as can be seen from Fig.~\ref{nikk4}.

As one can see, to a good approximation, the final $B-L$ asymmetry is actually fixed by the RHN number density at $a_*$, when RHNs produced from inflaton decays are kinematically shut off. Although $n_{B-L}$ may have had a more turbulent history, at $a > a_N$, we have the approximate simple relation 
\beq
n_{B-L} = \epsilon n_N(a_*) \left(\frac{a_*}{a} \right)^3 \,,
\eeq which can be directly visualized in Fig.~\ref{nikk4}, using $a_*/a_{\rm end}\simeq15$.

To determine the final baryon asymmetry $Y_B$, we also need to track the evolution of the temperature, or the radiation density $\rho_{\rm R}$. In Appendix \ref{sec:frag}, we provide analytic formulas for the radiation density in the absence of the contribution from RHN decays. However, because the branching ratio for inflaton decays to RHNs is of course non-zero in our case, these contributions must also be considered. In the following, we take into account the RHN contribution to the radiation.  

 For a heavy RHN,  $M_N\sim \mathcal{O}(10^{12})$~GeV, $a_*<a_\beta$, and the production of RHNs from inflaton decay is kinematically forbidden before the fragmentation of the inflaton condensate occurs.
As the inflaton is mostly in the form of  a condensate for  $a_{\rm end}<a<a_\beta$, which also  largely dominates the total energy density, the evolution of $\rho_{\rm R}$ is given by $\rho_{\rm R}\propto a^{-3}$ \cite{GKMO2}. For $a > a_\beta$,   as the inflaton (particle) decay term redshifts much faster than the RHN decay term, it  is possible that the RHN   becomes the main source of radiation production, thus modifying the temperature evolution. More precisely, we must compare the following two terms for $a_\beta<a<a_N$:
\begin{equation}
    \begin{aligned}
        \text{Inflaton decay: }\quad & \rho_{\delta \phi} \Gamma_{\delta \phi }~\simeq ~ c_4 \sqrt{\xi_0 \beta^b} {\hat m}^2 \rho_{\rm end}^\frac34 \left(\frac{\aend}{a} \right)^{5+\frac{b}{2}} \simeq 1.2 \times 10^{76}~\text{(GeV)}^4 y_{\phi ff}^2 \left(\frac{\aend}{a}\right)^{5.65} \, ,
        \\\text{RHN decay: }\quad & \rho_N\Gamma_N\simeq \frac{M_N^2y_N^2}{8\pi}n_{N}(a_*)\left(\frac{a_*} {a}\right)^3\\ & \simeq 2.6\times 10^{71}~\text{(GeV)}^4y_N^2 y_{\phi NN}^2 \left(\frac{\aend}{a}\right)^3 \frac{M_N}{10^{12}~\text{GeV}}\left(15-\frac{M_N}{10^{12}~\text{GeV}}\right) \, ,
    \end{aligned}
\end{equation}
where we used $a_* \simeq 15 \aend (10^{12}~{\rm GeV}/M_N)$.
We can then solve for the scale factor when the above two terms are equal. As long as this scale factor is larger than $a_N$, inflaton decay always dominates the production of radiation, and the expressions in Appendix \ref{sec:frag} still hold. In terms of the Yukawa couplings, this condition translates into:
\begin{equation}
    y_{\phi ff}\gtrsim \frac{y_{\phi NN}}{y_N^{0.325}}\frac{0.2\sqrt{15-(M_N/10^{12}~\text{GeV})}}{(M_N/10^{12}~\text{GeV})^{0.1625}} \, .
\end{equation}
For a choice of parameters which satisfy this relation, we show the numerical evolution of the radiation density in Fig.~\ref{rhornew} (blue curve). The asymptotic behaviors agree with the analytic formulas in Appendix \ref{sec:frag}. The radiation density in Fig.~\ref{rhornew} is compared to the inflaton energy density (green line) which is dominated by the condensate for $a < a_\beta$ and by inflaton quanta for $a > a_\beta$. The total energy density drops as $a^{-4}$. We also compare these to the mass density of RHNs (red curve). The solid curve shows the region where RHNs are non-relativistic, while along the dashed part 
RHN neutrinos are relativistic and $M_N n_N$ is only an approximation for $\rho_N$.
The drop-off in $\rho_N$ due to $N$-decay
is clearly seen at $a_N \simeq 10^7 \aend$. We note that while we derived the condition above needed for inflaton decays to always dominate the production of radiation, this is not required for our leptogenesis mechanism to be successful.

\begin{figure}[h!] \includegraphics[width=.7\textwidth]{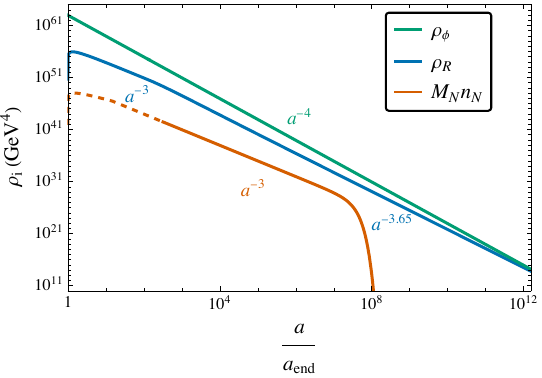} 
    \caption{The energy densities of $\phi$, radiation and RHNs as functions of $a/\aend$ with the asymptotic behaviors of $\rho_{\rm R}$ as labeled. For $M_N n_N$, the dashed line corresponds to $T>M_N$ and the solid line satisfies $T<M_N$ hence $\rho_N\simeq M_Nn_N$. We have chosen $\epsilon=-10^{-3}$, $y_{\phi ff}=0.002$, $y_N=y_{\phi NN}=10^{-6}$, $M_N=10^{12}$~GeV. }
    \label{rhornew}
\end{figure} 

\begin{figure}[h!] \includegraphics[width=.7\textwidth]{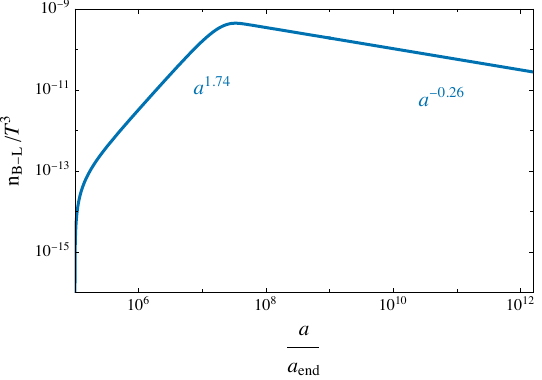} 
    \caption{$n_{B-L}/T^3$   terms of $a$ with the asymptotic behaviors indicated.  We have chosen $ \epsilon =-10^{-3}$, $y_{\phi ff}=0.002$, $y_N=y_{\phi NN}=10^{-6}$, $M_N=10^{12}$~GeV.}
    \label{nblt3}
\end{figure}

The evolution of the baryon asymmetry, $n_{B-L}/T^3$, is shown in Fig.~\ref{nblt3}. The rapid rise in $n_{B-L}$ occurs as $B-L$ is driven to positive values when inverse decays are kinematically suppressed. The slopes seen in this figure can be understood as follows. From Eq.~(\ref{nblbiga}), when $N$ is nonrelativistic and at suitably large values of $a$ (so as to erase the effect of the initial conditions set by inverse decays), $n_{B-L} \propto a^{-1}$. But from Eq.~(\ref{rhorfrag2}), 
we see that $T^4 \propto a^{3+\frac{b}{2}}$ and for $b=1.3$, we obtain the scaling $n_{B-L}/T^3 \propto a^{1.74}$. Similarly, at later times ($a > a_N$), when $n_{B-L} \propto a^{-3}$, we have that $n_{B-L}/T^3 \propto a^{-0.26}$. 

We are now in a position to derive an analytic expression for the final baryon asymmetry for $k=4$. As argued above, we can evaluate $n_{N}$ first at $a_*$, and then obtain the final baryon asymmetry at $\arh$ with
\begin{equation}
    n_{B}(a_{\rm RH})=\frac{8}{23}\epsilon n_{N}(a_*)\left(\frac{a_*}{\arh}\right)^{3} \, .
    \label{nrhn*}
\end{equation}
We can then use $Y_{B}(a_{\rm RH})=\frac{n_B(a_{\rm RH})}{s(a_{\rm RH})}$ to evaluate the final baryon asymmetry. It is interesting to compare this result with what we found for $k=2$. Recall that evaluating the relevant quantities at $a_{\rm sph}$ was important for $k=2$, as 
RHNs continued to be produced (and decay producing a lepton asymmetry) continuously from $\aend$ to $\arh$. However, for $a > a_{\rm sph}$, any new contribution to the lepton asymmetry cannot be processed into a baryon asymmetry as the sphaleron transitions drop out of equilibrium.  This led to the the severe dilution of the asymmetry when $\arh > a_{\rm sph}$. This picture is significantly altered for $k=4$ as the production of RHNs from inflaton decay
stops at $a_*$. Indeed, in this case, we can obtain the baryon asymmetry at $\arh$ directly from $n_N(a_*)$ using Eq.~(\ref{nrhn*}), without evaluating any quantities directly at $a_{\rm sph}$. The result is the following:
\begin{align}
Y_B(a_{\rm RH}) &=\frac{n_B(a_{\rm RH})}{s(a_{\rm RH})}=
\frac{8}{23}\frac{\epsilon n_N(a_*)\left(\frac{a_*}{a_{\rm end}}\right)^3 \left(\frac{a_{\rm end}}{a_{\rm RH}}\right)^3}{s(a_{\rm RH})} = \frac{8}{23}\frac{\epsilon n_N(a_*)\left(\frac{a_*}{a_{\rm end}}\right)^3 \alpha^{3/4}T_{\rm RH}^3 \rho_{\rm end}^{-3/4}}{\frac{2 \pi^2 g_*(a_{\rm RH})}{45}T_{\rm RH}^3} \label{Eq:YbaRH0},
\end{align} 
where we have left the factors of $T_{\rm RH}$ explicit in the numerator and denominator so that the origin of the cancellation and the resulting $T_{\rm RH}$-independence is apparent. The fact that the final asymmetry in Eq.~(\ref{Eq:YbaRH0}) is independent of $T_{\rm RH}$ up to factors of $g_*(T_{\rm RH})$ is somewhat surprising, and will be crucially significant for our final results.

Notice that Eq.~(\ref{Eq:YbaRH0}) does not contain a simple dilution factor of the form $(T_{\rm RH}/T_{\rm sph})^{\gamma}$, which was our original expectation as shown in Eq.~(\ref{dilute}). Indeed, in the discussion surrounding Eq.~(\ref{dilute}), we argued that the final asymmetry could be obtained from $Y_B(a_{\rm sph})$ suitably diluted to $Y_B(\arh)$ which then remains constant for $a>\arh$. With $n_B \propto a^{-3}$ for $a > a_{\rm sph}$ and $T \propto a^{-(3k-3)/(2k+4)}$, we would we expect a dilution factor, $(\Trh/T_{\rm sph})^\gamma$ with $\gamma = (7-k)/(k-1)$. For $k=2$, this gives $\gamma = 5$. For $k=4$, $\gamma = 1$, and naively one might assume that the asymmetry is only diluted by a factor of $\Trh/T_{\rm sph}$. However for $k=4$, there is a key difference in the evolution of the asymmetry. 
RHN neutrinos are only produced by inflaton decay when $m_{\rm eff}>2M_N$. These decays become kinematically forbidden at $a_*$. Thus, unlike the case for $k=2$ where decays to $N$ are important all the way to $T_{\rm sph}$, for $k=4$, we must simply dilute from $a_*$ down to $\arh$. While this may seem to be an extreme amount of tuning, we draw attention to the fact that for $k=2$, we are fixing the temperature scale as a cutoff ($T_{\rm sph}$), while for $k=4$,
it is $a_*$, which is independent of $T_{\rm max}$ and $\Trh$.  Let us follow through with the argument: If we dilute the asymmetry (with $\gamma = 1$)  from $T(a_*)$
to $\Trh$, we would expect \footnote{Note that $Y_B(a_*)$ in Eq.~(\ref{Eq:Ybk4fermion}) is an inferred quantity corresponding to a future baryon asymmetry which will be generated by an existing RHN population, since at $a=a_*$ the RHNs will not yet have decayed.}
\beq
Y_B(\arh) = Y_B(a_*) \left( \frac{\Trh}{T_*} \right) = \frac{8}{23}\frac{\epsilon n_N(a_*)}{s(a_*)}  \left( \frac{\Trh}{T_*} \right) \, .
\label{Eq:Ybk4fermion}
\eeq
Since $a_*$ is determined by the fixed quantities,
$M_N$ and ${\hat m}$, we can define
$T_*$ in terms of $a_*$, $\aend$, and $T_{\rm max}$, namely $T_* = T_{\rm max}(\aend/a_*)^{3/4}$. 
From Eqs.~(\ref{tmax4}) and (\ref{trh4nofrag}), 
we see that $T_{\max} \propto \Trh^\frac14$. 
Note that entropy is {\em not} conserved during the entire reheating process, and $s$ is {\em not} proportional to $a^{-3}$ as it would in an adiabatically expanding universe. Rather, $s(a_*) \propto T_*^3$, and hence the combination $\Trh/(s(a_*)T_*) \propto \Trh/T_*^4 \propto \Trh/T_{\rm max}^4$ is constant and independent of $\Trh$ as argued in Eq.~(\ref{Eq:YbaRH0}).

Thus to determine the baryon asymmetry at reheating, we need only $n_N(a_*)$ which is easily obtained from Eq.~(\ref{solnnainf}) and Eq.~(\ref{Eq:aStar}) such that
\beq
n_N(a_*)\simeq \frac{2 \sqrt{\rho_{\rm end}} M_P y_{\phi NN}^2}{\sqrt{3} \pi}\left(\frac{M_N}{\hat{m}}\right)^2 \, .
\eeq
Applying this result to Eq.~(\ref{Eq:YbaRH0}) and again using Eq.~(\ref{Eq:aStar}) we obtain 
\beq
Y_B(a_{\rm RH})\simeq \frac{270^\frac14}{23}\frac{|\epsilon| y_{\phi NN}^2}{ \pi^{3/2} g_*(a_{\rm RH})^{1/4}}\left( \frac{M_P}{\rho_{\rm end}^{1/4}}\right)\left(\frac{\hat{m}}{2 M_N}\right) \, ,
\label{YBarh}
\eeq
where we have ignored the lower limit of integration in this approximation. 
Note that since $\epsilon \propto M_N$ (see Eq.~(\ref{Eq:epsilon})), we expect the final asymmetry to be effectively independent of $M_N$ in addition to being independent of $\Trh$.

We are now in a position to determine whether the observed baryon asymmetry of $Y_B = 8.7 \times 10^{-11}$ is compatible with low reheating temperatures.  The details of our leptogenesis process depend upon the following set of parameters: $y_{\phi ff}, \epsilon,y_{\phi NN},M_N$ and $y_N$. Alternatively we can substitute $\Trh$ for $y_{\phi ff}$ and $\delta_{\rm eff}$ for $\epsilon$. However, as we have seen above, not all of these parameters have a direct impact on the baryon asymmetry. While these parameters are adjustable, they are subject to various basic constraints such as the BBN bound on $T_{\rm RH}$ and $\delta_{\rm eff} \lesssim 1$. In contrast, we consider $\rho_{\phi}(a_{\rm end})$ to be fixed for a given inflaton potential and the Planck normalization of the CMB anisotropy spectrum. For simplicity we have fixed $m_{\nu_i}=0.05$~eV. Of this set of five parameters, we found that the final baryon asymmetry is largely insensitive to three of them, namely $T_{\rm RH},M_N$ and $y_N$, so long as the following hierarchy exists: $a_*<a_{N}<a_{\rm sph},a_{\rm RH}$ \footnote{At the end of this section below, we derive precise conditions on the permissible values of $y_N$.}. Stated another way, this hierarchy implies that kinematic shutoff of the $\phi \rightarrow NN$ channel occurs prior to RHN decay (this is practically guaranteed for high $M_N\simeq 10^{12}$~GeV) which in turn occurs prior to the sphaleron transition and reheating. Indeed, $y_N$ can be freely adjusted over a large range without impacting the final baryon asymmetry as long as the above hierarchy of scales is satisfied and the RHN energy density does not overtake the inflaton's energy density, which would disrupt the reheating process. While the final asymmetry is largely insensitive to $T_{\rm RH}$, there is a mild sensitivity to $T_{\rm RH}$ through the number of relativistic degrees of freedom as can be seen from Eq.~(\ref{YBarh}).

\begin{figure}[h!] \includegraphics[width=.9\textwidth]{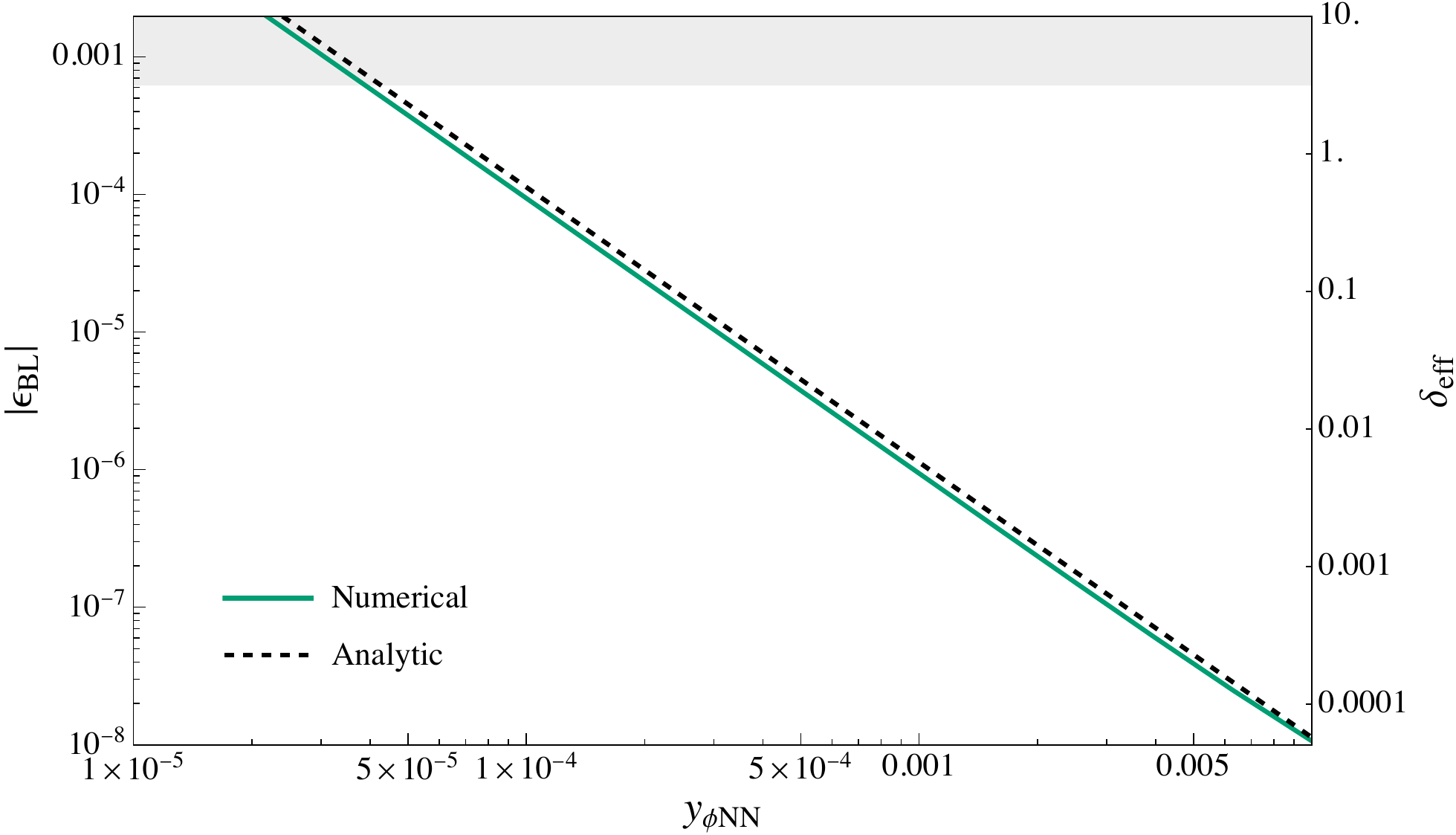} 
    \caption{Parameter space in the ($y_{\phi NN},\epsilon$) plane consistent with the correct baryon asymmetry. We have taken $T_{\rm RH}=1$~GeV, $y_N=10^{-2}$, $M_N=10^{12}$~GeV, and $m_{\nu,i}=0.05$~eV. The gray shaded region corresponds to $\delta_{\rm eff}>\pi$. The black dashed line used for the analytic estimate corresponds to Eqs.~(\ref{Eq:epsAnalytic}) and(\ref{deltaeff}).
    }
    \label{Fig:epsyPhiNN}
\end{figure}

To capture the viable parameter space consistent with the observed asymmetry, we can use our analytic results above to obtain an expression which relates $y_{\phi NN}$ and $|\epsilon|$. For example, we can invert Eq.~(\ref{YBarh}), to determine $y_{\phi NN}$ terms of $M_N$ and $\epsilon$, 
\begin{align}
y_{\phi NN}&=(Y_{\rm B,obs})^{1/2}\left(\frac{23 \pi^\frac32}{270^\frac14} \right)^\frac12 \left(\frac{\rho_{\rm end}^{\frac14}}{M_P}\right)^{\frac12}\left(\frac{\hat{m}}{2M_N}\right)^{-\frac12}|\epsilon|^{-\frac12}g_{*}(a_{\rm RH})^{\frac18}  \nonumber \\
&=1.1 \times10^{-3}\left(\frac{10^{-6}}{|\epsilon|}\right)^{\frac12} 
\left(\frac{\hat{m}}{3.0\times10^{13}\text{ GeV}}\frac{10^{12}\text{ GeV}}{M_N} \right)^{-\frac12}
\left(\frac{4 g_{*}(a_{\rm RH})}{247}\right)^{\frac18} \, ,
\label{ynnep}
\end{align} 
or equivalently
\begin{align}
|\epsilon|&=(Y_{\rm B,obs})\left(\frac{23 \pi^{\frac32}}{270^\frac14} \right)\left(\frac{\rho_{\rm end}^{\frac14}}{M_P}\right)\left(\frac{\hat{m}}{2M_N}\right)^{-1}y_{\phi NN}^{-2}g_{*}(a_{\rm RH})^{\frac14} \nonumber\\
&\simeq 10^{-6}\left(\frac{1.1\times10^{-3}}{y_{\phi NN}}\right)^2\left( \frac{\hat{m}}{3.0\times10^{13}\text{ GeV}}\frac{10^{12}\text{ GeV}}{M_N}  \right)^{-1} \left(\frac{4 g_{*}(a_{\rm RH})}{247}\right)^{1/4} \, ,
\label{Eq:epsAnalytic}
\end{align} 
where in the last line we have used $Y_{\rm B,obs}=8.7\times10^{-11}$ and we have taken $g_*(T_{\rm RH}=1 \text{ GeV})=247/4$ to facilitate comparison with our numerical results in Fig.~\ref{Fig:epsyPhiNN}.
Then, substituting Eq.~(\ref{Eq:epsilon}) and the relevant type-I seesaw relations into Eq~(\ref{ynnep}), we find 
\bea
y_{\phi NN}
&= 7.6 \times 10^{-4} \left(\frac{10^{-2}}{\delta_{\rm eff}}\right)^{1/2}\left(\frac{0.05 \text{ eV}}{m_{\nu_i}}\right)^{1/2}\left(\frac{4 g_*(T_{\rm RH})}{247}\right)^{1/8} \label{Eq:yPhiNN2},
\eea where we have used the convention $v\simeq 174\,{\rm GeV}$, so that
$m_{\nu_i}\simeq y_i^2 v^2/2M_i$. Notice that the mass of the lightest RHN, $M_N$, has dropped out as expected. The solution for $\delta_{\rm eff}$ becomes
\beq
\delta_{\rm eff}=10^{-2}\left(\frac{7.6\times 10^{-4}}{y_{\phi NN}}\right)^2\left(\frac{0.05 \text{ eV}}{m_{\nu_i}}\right)\left(\frac{4 g_{*}(a_{\rm RH})}{247}\right)^{1/4} \, .
\label{deltaeff}
\eeq
Eqs.~(\ref{ynnep})-(\ref{deltaeff}) are key analytic results which summarize the available parameter space consistent with the correct baryon asymmetry for our model when $k=4$. A significant consequence of the above expressions is that the correct baryon asymmetry is largely independent of $T_{\rm RH}$ and $M_N$.
We have thus obtained the important result that reheating temperatures as low as the BBN bound are allowed with this mechanism, and we confirm this in our numerical results below. Our final available parameter space is sensitive primarily to $\epsilon$ and $y_{\phi NN}$.

Consequently, the parameter space consistent with the correct baryon asymmetry for $k=4$ can be displayed in the ($y_{\phi NN},\epsilon$) plane as shown in Fig.~\ref{Fig:epsyPhiNN}. On the left-hand axis, we show $|\epsilon|$, while the right hand axis shows the corresponding values of $\delta_{\rm eff}$ according to Eq.~(\ref{Eq:epsilon}).  The numerical results agree remarkably well with our analytic result in Eqs.~(\ref{Eq:epsAnalytic}) and (\ref{deltaeff}), with $|\epsilon|\propto y_{\phi NN}^{-2}$,  for fixed $M_N$. The fact that $\epsilon$ and $y_{\phi NN}$ are inversely correlated can be easily understood. Specifically, if we increase $y_{\phi NN}$, there will be a greater maximum number density of RHNs prior to kinematic shutoff of the $\phi \rightarrow NN$ channel. This in turn means that we do not need as large a value of $\epsilon$ to produce the required asymmetry. 
As one can see, from the requirement that $\delta_{\rm eff}\leq \pi$, we find a lower limit of $y_{\phi NN} \gtrsim 4.3 \times 10^{-5}$ and an upper limit of $|\epsilon|\lesssim 6\times 10^{-4}$. The small discrepancy between the numerical and analytic results can be attributed to the fact that our analytic estimate neglected the kinematic factor in the decay rate $\Gamma_{\phi\rightarrow NN}$ for simplicity, and because we dropped the term produced by the lower limit of integration in our computation of $n_N(a_*)$.

\begin{figure}[h!] \includegraphics[width=.9\textwidth]{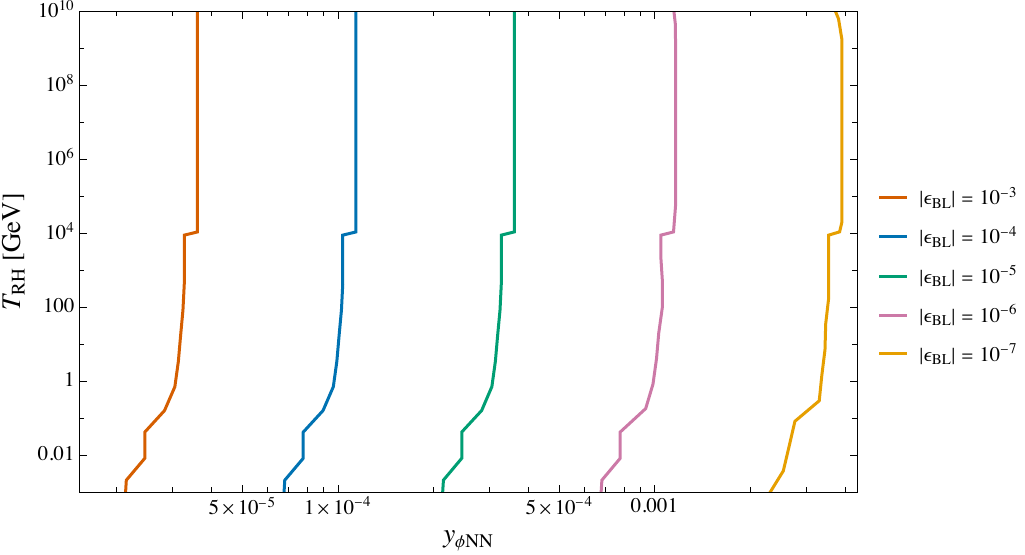} 
    \caption{Parameter space in the ($y_{\phi NN},T_{\rm RH}$) plane consistent with the correct baryon asymmetry. We have taken $y_N=10^{-4}$, $M_N=10^{12}$~GeV, and $m_{\nu,i}=0.05$~eV.}
    \label{Fig:TRHyPhiNN}
\end{figure}

The result in Fig.~\ref{Fig:epsyPhiNN} is largely independent of the reheating temperature, and would look nearly identical for $T_{\rm RH}=4$~MeV or $T_{\rm RH}=100$~GeV.  There is only a very mild dependence on $\Trh$ due to the $1/4$ power of the number of relativistic degrees of freedom, $g_*(T_{\rm RH})$ at reheating, which can be seen in Eq.~(\ref{Eq:epsAnalytic}).
We further highlight the $T_{\rm RH}$-independence by depicting the results in the $(y_{\phi NN},T_{\rm RH})$ plane in Fig.~\ref{Fig:TRHyPhiNN} for several choices of $|\epsilon|$. The primary takeaway from Fig.~\ref{Fig:TRHyPhiNN} is that the pair of $(y_{\phi NN},|\epsilon|)$ values required to obtain the correct baryon asymmetry does not change appreciably even over 13 orders of magnitude in $T_{\rm RH}$. For instance, one can see that for a fixed choice of $|\epsilon|$, $y_{\phi NN}$ does not even change by a factor of 2 for $T_{\rm RH}=10^{10}$~GeV compared to $T_{\rm RH}=4$~MeV. The subtle changes that are observed are explained by the dependence of the baryon asymmetry on $g_{*}(T_{\rm RH})^{1/4}$, for instance as 
seen in Eq.~(\ref{YBarh}). 
For example, the shift at $\Trh = 10$~TeV is due to the inclusion of all of the MSSM superpartners. 

Before concluding this section, we lastly provide a more detailed discussion of the role of the RHN Yukawa coupling, $y_N$. Notice that none of our key analytic results above contain $y_N$. Indeed, the final baryon asymmetry is independent of $y_N$ given some basic constraints which we will now derive. Recall that kinematic shutoff of the non-thermal production channel, $\phi \rightarrow NN$, occurs at $a=a_*$, after which the RHN number density simply dilutes as $n_N(a>a_*)\propto a^{-3}$. This dilution will continue until the RHNs begin to decay efficiently. The scale factor at which RHN decays become efficient is entirely determined by $y_N$ for a fixed choice of $M_N$. Now, because the RHN energy density after $a_*$ evolves as $\rho_N \simeq M_Nn_N(a_*)(a_*/a)^{3}$, and the inflaton energy density evolves as $\rho_\phi(a)\propto a^{-4}$, the RHN density will eventually overtake the inflaton energy density if the RHNs do not decay quickly enough. Note that we cannot allow the RHN density to overtake the inflaton density and still achieve successful standard leptogenesis with low reheating temperatures. \footnote{As we mentioned in Section \ref{k2}, if the RHN density overtakes the inflaton density, then reheating will occur via RHN decays to SM particles. For low reheating temperatures, RHN decays would therefore occur primarily after sphaleron decoupling (corresponding to very small $y_N$) and would result in a lepton asymmetry that would not be processed to a baryon asymmetry.} Therefore, while $y_N$ does not directly alter the baryon asymmetry, we do require the basic constraint that $y_N$ is sufficiently large to prevent the RHN density from overtaking the inflaton density. We can derive an estimate of this constraint as follows. First, we can find the scale factor, $a_{\rm eq}/a_{\rm end}$, at which the RHN density would equal the inflaton energy density if RHN decay occurs sufficiently late. This can be found by setting the RHN energy density for $a>a_*$ equal to the inflaton density, which gives
\beq
M_N n_N(a_*)\left(\frac{a_*}{a_{\rm eq}}\right)^3=\rho_{\rm end}\left(\frac{a_{\rm end}}{a_{\rm eq}}\right)^4 \, ,
\eeq
so that
\beq
\frac{a_{\rm eq}}{a_{\rm end}}=4\sqrt{3}\pi\frac{\sqrt{\rho_{\rm end}}}{M_P \hat{m}} y_{\phi NN}^{-2} \, , 
\eeq 
where we have used Eqs.~(\ref{Eq:aStar}) and (\ref{solnnainf}). Having obtained the scale factor at which RHN-inflaton equality would occur, we can now require that RHN decay begins efficiently prior to this scale factor. Namely, we can require $\Gamma_N > H(a_{\rm eq})$ which implies
\beq
\frac{y_{N}^2 M_N}{8\pi} > \frac{\sqrt{\rho_{\rm end}}}{\sqrt{3} M_P}\left(\frac{a_{\rm end}}{a_{\rm eq}}\right)^2 \, ,
\eeq
or
\beq
y_N>\left(\frac{M_P}{M_N}\right)^{1/2}\left(\frac{\hat{m}}{\rho_{\rm end}^{1/4}}\right)(6\sqrt{3}\pi)^{-1/2}y_{\phi NN}^2 \simeq 1.5 y_{\phi NN}^2 \left(\frac{10^{12} \text{ GeV}}{M_N}\right)^{1/2} \, .
\eeq 
This provides a constraint on the minimum of $y_N$, but there is also a constraint on the maximum permissible value of $y_N$. Specifically, in order for our analytic estimates of the baryon asymmetry above to be valid, we require that RHNs do not begin decaying efficiently until after kinematic shutoff of the $\phi \rightarrow NN$ channel. We can therefore derive the following constraint from $\Gamma_N < H(a_{*})$ which gives
\beq
\frac{y_{N}^2 M_N}{8\pi} < \frac{\sqrt{\rho_{\rm end}}}{\sqrt{3} M_P}\left(\frac{a_{\rm end}}{a_{*}}\right)^2 \, ,
\eeq
or
\beq
y_N <\left(\frac{\sqrt{\rho_{\rm end}}M_N}{M_P \hat{m}^2}\right)^{1/2}\left(\frac{32\pi}{\sqrt{3}}\right)^{1/2}\simeq 0.9\left(\frac{M_N}{10^{12}\text{ GeV}}\right)^{1/2} \, .
\eeq
Combining the two constraints, we have the following condition on $y_N$ for our intended mechanism to robustly hold:
\beq
1.5 y_{\phi NN}^2 \left(\frac{10^{12}\text{ GeV}}{M_N}\right)^{1/2} < y_N < 0.9\left(\frac{M_N}{10^{12}\text{ GeV}}\right)^{1/2}.
\label{Eq:yNconstraint}
\eeq  These are not very restrictive constraints, and it is straightforward to check that our parameters used in Figs.~\ref{Fig:epsyPhiNN} and \ref{Fig:TRHyPhiNN}, and throughout this section satisfy the above conditions.

\section{Gravitational production of radiation and RHNs}
\label{gravprod}

In addition to the terms included in the coupled Boltzmann equations in Eq.~(\ref{Eq:Boltz}), there will also be unavoidable contributions to $\rho_{N}$ and $\rho_{\rm R}$ from the gravitational scattering of the inflaton \cite{Clery:2021bwz}, namely $\phi \phi \rightarrow h_{\mu \nu} \rightarrow \text{SM SM}$ and $\phi \phi \rightarrow h_{\mu \nu} \rightarrow N N$. In the following, we explain why these additional sources are negligible in this context. For example, the source term from gravitational inflaton scattering on the RHS of the Boltzmann equation for $\rho_{\rm R}$ is of the form
\begin{equation}
N \frac{\rho_{\phi}^2 \omega}{16 \pi M_{P}^4}\Sigma^h_k,
\end{equation} where $N=98$ is the number of scalar degrees of freedom in the MSSM, $\omega$ is the oscillation frequency of the inflaton condensate and $\Sigma^h_k$ is a numerical factor resulting from summing over Fourier modes of the oscillating inflaton condensate. 

For minimal gravity, gravitational inflaton scattering alone cannot lead to reheating of the universe that respects the BBN bound ($T_{\rm RH}>4$~MeV) unless $k\ge 10$ \cite{Co:2022bgh,Barman:2022qgt}. If the Higgs boson couples to curvature \cite{Clery:2022wib}, the lower limit on $k$ can be relaxed to $k\ge 6$.  Similarly the gravitational production of RHNs leading to successful leptogenesis requires $k\ge 6$.  However, the gravitational production of radiation may modify the temperature evolution of the bath. For the parameter values we have used in this work, the gravitational contribution to the radiation bath will be negligible for $k\geq4$ (recall that these are the cases of greatest interest for our work) since the $T_{\rm max}$ produced via inflaton decays will exceed the corresponding quantity from gravitational scattering. For $k=2$, the gravitational contribution to the radiation bath may transiently dominate early in the initial phase of the reheating period. The gravitational contribution for $k=2$ will lead to an increase in $T_{\rm max}$ over the naive value, and the initial evolution of the radiation bath will be $\rho_{\rm R}\propto a^{-4}$ rather than $\rho_{\rm R}\propto a^{-3}$ during this initial period. However, subsequently, the gravitational contribution to the radiation density will be exceeded by the contribution from inflaton decay and will be negligible through the remainder of the reheating period. The net effect on the final baryon asymmetry is negligible. Recall however, that successful leptogenesis for $k=2$ is not possible. These effects are dependent on $y_{\phi ff}$ and consequently upon $T_{\rm RH}$. For sufficiently large $y_{\phi ff}$, the gravitational contribution to the radiation bath will always be negligible even for $k=2$. 

Next, we consider the gravitational scattering of the inflaton to RHNs. Due to a helicity suppression, the gravitational contribution does not alter the RHN evolution for any portion of our parameter space. From Ref. \cite{Clery:2021bwz}, we know that the rate for $\phi \phi \rightarrow h_{\mu \nu} \rightarrow NN$ scattering is given by 
\beq
R_{\rm grav}=\frac{\Sigma_{4}^{1/2}}{2\pi}\frac{\rho_{\phi}^2M_N^2}{M_P^4 m_{\phi}^2},
\label{Eq:gravRHN}
\eeq where the $\Sigma_4^{1/2}=0.061$ is a numerical factor produced by summing over Fourier modes of the inflaton potential. To account for this contribution, we would include Eq.~(\ref{Eq:gravRHN}) as a source term in the RHS of our Boltzmann equation for the RHN number density in Eq.~(\ref{Eq:Boltz}). To determine whether this gravitational scattering term will be relevant for our model, we can calculate the number density of RHNs produced by gravitational scattering vs. decays and then compare the two. It is suitable to integrate the respective number density contributions from $a_{\rm end}$ to $a_{*}$ for this purpose, since gravitational production peaks near $a_{\rm end}$ and decay production ceases after $a_*$. Doing this, we find the following for the gravitational contribution:
\beq
n_N^{\rm grav}(a_*)\simeq \frac{4 \sqrt{3} \Sigma_4^{1/2}}{\pi}\frac{\rho_{\rm end}^{3/2}}{M_P^3}\left(\frac{M_N}{\hat{m}}\right)^5.
\eeq 
Then we can compare this with the contribution from inflaton decays
\beq
n_N^{\rm dec}(a_*)\simeq \frac{2}{\sqrt{3}\pi}y_{\phi NN}^2\sqrt{\rho_{\rm end}}M_P\left(\frac{M_N}{\hat{m}}\right)^2.
\eeq 
We can then readily determine the condition on $y_{\phi NN}$ such that the gravitational contribution would dominate, namely
\beq
y_{\phi NN} < \left(\frac{6 \Sigma_4^{1/2}\rho_{\rm end}}{M_P^4}\right)^{1/2}\left(\frac{M_N}{\hat{m}}\right)^{3/2} \simeq y_{\phi NN}< 1.8\times10^{-8} \left(\frac{M_N}{10^{12}\text{ GeV}}\right)^{3/2}.
\eeq 
For this work, we always require $y_{\phi NN}\gg 10^{-8}$ to obtain the correct baryon asymmetry, so the gravitational contribution to the RHN density is always negligible. For instance, the plot of $\rho_N(a)$ shown in Fig.~\ref{rhornew} for $y_N = 10^{-6}$ and $y_{\phi NN} = 10^{-6}$ is unaltered when the gravitational contribution is included. In sum, gravitational scattering of the inflaton to both SM particles and RHNs can be safely neglected in all of our cases of interest.

\section{Scalar reheating and $k\geq 4$}
\label{other}

In Section \ref{k4}, we studied $k=4$ for fermionic reheating. For $k>4$, inflaton decays to fermions are strongly affected by fragmentation \cite{Garcia:2023dyf}. Reheating is no longer possible for values of the critical coupling which may lead to $\arh > a_\beta$. Therefore, for larger $k$, we consider only bosonic reheating via inflaton decays to scalars. 
For general $k$, the inflaton potential near the minimum is given by Eq.~(\ref{appk}). Taking  $k=6$ and $\Trh=1$~GeV, we then have the following associated parameters: $\rho_{\rm end}^{1/4}= 5.1\times 10^{15} $~GeV,   $\lambda=1.35\times 10^{-10}$, $n_s=0.967$, and $N_*=59.7$.
Similarly, for  $k=8$ and $\Trh=1$~GeV, we have $\rho_{\rm end}^{1/4}= 5.0\times 10^{15} $~GeV,  $\lambda= 1.28\times 10^{-10}$, $n_s=0.968$, and $N_*=61.6$.

Consider bosonic reheating via inflaton decays to scalars, namely $\phi\rightarrow bb$ with coupling $\mu \phi bb$. For general $k$, the decay rate is given by \cite{GKMO2}:
\begin{equation}
    \Gamma_{\phi\rightarrow bb}=\gamma_\phi \left(\frac{\rho_\phi}{M_P^4}\right)^{\frac{1}{k}-\frac{1}{2}
    },\quad\text{with}\quad \gamma_\phi=\frac{\mu^2}{8\pi \sqrt{k(k-1)}(2^{-2+\frac{k}{2}} 3^{1-\frac{k}{2}}\lambda )^{1/k}M_P} \, .
\end{equation}
The maximal temperature in terms of $\mu$ is given by
\begin{equation}
      T_{\rm max}=10^{12}\text{ GeV}\left(\frac{\mu}{\mu_k}\right)^{1/2} \, ,
\end{equation}
with $\mu_k=6.0\times 10^7$~GeV for $k=4$, $\mu_k=6.5\times 10^7$~GeV for $k=6$, and $\mu_k=7.0\times 10^7$~GeV for $k=8$. For the low reheating temperatures that we consider with $\Trh<T_{\rm sph}$, we have $\mu\ll 1$~GeV. Therefore for heavy RHN, $T_{\rm max}\ll M_N\sim \mathcal{O}(10^{12})$~GeV, and the thermal production can be completely neglected. 
The reheating temperature for decays to scalars for $k=2$ and $k=4$ was given in Eqs.~(\ref{trhs2}) and (\ref{trhs4}). The general expression for $\Trh$ (in the absence of fragmentation) is
\beq
\alpha_{\rm RH} \Trh^4 = 3\cdot 2^\frac{4-7k}{2k-2} \left(\frac{k}{(k-1)(2k+1)^2}\right)^{\frac{k}{2k-2}} \lambda^\frac{-1}{k-1} \left(\frac{\mu^2 M_P^\frac{2(k-2)}{k} }{\pi} \right)^\frac{k}{k-1} \, .
\eeq
The result for $\Trh$ including fragmentation is similar \cite{Garcia:2023dyf}. 

For $k\geq4$, the RHN production from inflaton decay ceases at \begin{equation}
    \frac{a_*}{\aend}=\left(\frac{{\hat m}}{2M_N}\right)^{\frac{k+2}{3(k-2)}} \, .
\end{equation}
 Note also that the values of ${\hat m}$ are similar for the larger values of $k$, e.g.  $\hat{m}\simeq 3.0\times 10^{13}$ GeV
 for $k=6$ and $k=8$.

  For $a<a_*$, the analytic solution for $n_N$ is
  \begin{equation}
    n_N(a<a_*)=\frac{\sqrt{3 \rho_{\rm end}} M_P y_{\phi NN}^2k}{24 \pi } \left( \frac{\aend}{a} \right)^\frac{3k}{k+2} \left( 1-\left(\frac{\aend}{a}\right)^\frac{6}{k+2} \right) \, ,
    \label{solnnainfk}
\end{equation}
which reduced to Eq.~(\ref{solnnainf}) for $k=4$.
After the kinematic suppression of the inflaton decay, using the non-relativistic limit of the decay rate \eqref{gammansmallt}, the solution for $n_N$ is \begin{equation}
     n_N(a>a_*)=n_N(a_*)\left(\frac{a_*}{a}\right)^3\exp\left[ \frac{(k+2)M_Ny_N^2}{24k\pi H_{\rm end}\aend^{\frac{3k}{k+2}}}\left(a_*^{\frac{3k}{k+2}}-a^{\frac{3k}{k+2}}\right)\right]
\end{equation}
As in the case for $k=4$, $n_N$ first redshifts as $a^{-3}$ when $a>a_*$, and then the RHNs decay away around $a_N$, with
\begin{equation}
    \frac{a_N}{\aend}=\left( \frac{24k\pi H_{\rm end}}{(k+2)M_Ny_N^2}   \right)^{\frac{k+2}{3k}} \, .
\end{equation}

For $n_{B-L}$,  Eq.~\eqref{nblequation} yields:
\begin{equation}
    n_{B-L}(a<a_*)=-\frac{\epsilon M_N M_P^2 y_N^2y_{\phi NN}^2 k}{64 \pi^2 } \left[ \frac13\left( \frac{a}{\aend}\right)^3 - \frac{k+2}{3k} \left( \frac{a}{\aend} \right)^2 + \frac{2}{3k} \right] \left(\frac{\aend}{a} \right)^3 \, ,\label{nblsmallak}
\end{equation}
and
\begin{equation}
       n_{B-L}(a>a_*)=\frac{a_*^3}{a^3}\left[  n_{B-L}(a_*)+\epsilon n_N(a_*) \exp \left[ \frac{(k+2) \left(a_*^{\frac{3k}{k+2}}-a^{\frac{3k}{k+2}}\right) }{a_N^{\frac{3k}{k+2}}}\right]-\epsilon n_N(a_*)  \right] \, .
\end{equation}
Soon after $a_*$, $n_{B-L}\propto a^{-\frac{6}{k+2}}$ and for $a>a_N\gg a_*$, we have $n_{B-L}\propto a^{-3}$.
These solutions for $n_N$ and $n_{B-L}$ for general $k$ reduce to those given in Sec.~\ref{k4} when  $k=4$.  Note that when $a <a_\beta$, the evolutions of $n_N$ and $n_{B-L}$ are independent of whether the inflaton decays to fermions or scalars.

After $a_\beta$, the fragmentation process for scalar reheating differs from the fermionic reheating case in Sec.~\ref{k4}. The decay rate of the inflaton particles is given by \cite{Garcia:2023dyf}:
\begin{equation}
\Gamma_{\delta\phi}
=
\frac{\mu^2}{8\pi \bar{E}}
=
\frac{\mu^2}{8\pi c_e M_P^2}
\left(
\frac{\rho_{\rm end}}{M_P^4}
\right)^{-1/4}
\beta^{\frac{k-4}{2(k+2)}}
\left(
\frac{a}{a_{\rm end}}
\right)
M_P \, ,
\end{equation}
where $\beta \equiv a_{\beta}/\aend$. Note that this decay rate has no dependence on $\xi$. This leads to the reheating temperature:
\begin{equation}
 \Trh=
\left(
\frac{30}{g_\rho \pi^2}
\right)^{1/4}
M_P
\left(
\frac{\sqrt{3}\,\mu^2}{8\pi c_e M_P^2}
\right)^{1/3}.
\end{equation}
which is independent of $k$, $\xi$ and $a_\beta$.  $\arh$ is given by
\begin{equation}
\frac{a_{\rm RH}}{a_{\rm end}}
=
\left(
\frac{\sqrt{3}\,\mu^2}{8\pi c_e M_P^2}
\right)^{-1/3}
\left(
\frac{\rho_{\rm end}}{M_P^4}
\right)^{1/4}
\beta^{-\frac{k-4}{2(k+2)}} .
\end{equation}
Finally, from Eq.~(\ref{Eq:YbaRH0}), we obtain the baryon asymmetry:
\beq
Y_{B}(a_{\rm RH}) 
\simeq 
\frac{8 |\epsilon|  n_N(a_*)
\left(\frac{a_*}{a_{\rm end}}\right)^3 \left(\frac{a_{\rm end}}{a_{\rm RH}}\right)^3}
{23 s(a_{\rm RH})}
\propto  \frac{|\epsilon| n_N(a_*)\left(\frac{a_*}{\aend}\right)^3}{\left(\beta^{\frac{2(4-k)}{k+2}}\rho_{\rm end}\right)^{\frac{3}{4}}} \, .
\label{Eq:Ybscalars}
\eeq 

As can be seen from Eq.~(\ref{Eq:Ybscalars}), the final baryon asymmetry is once again simply related to the size of a nonthermally generated early population of RHNs ($n_N(a_*)$) and the CP-violating parameter $|\epsilon|$. Interestingly, the above expression is again independent of $\Trh$. Thus, we have found that the general features of leptogenesis for $k=4$ and fermionic reheating that we investigated in Section 5 are also present for bosonic reheating with scalar final states, as can be seen from the similarities between Eqs.~(\ref{Eq:Ybscalars}) and (\ref{Eq:YbaRH0}). These results demonstrate that standard leptogenesis is a robust mechanism for producing the baryon asymmetry of the universe for arbitrarily low reheating temperatures above the BBN bound when the equation of state of the inflaton condensate is $w_{\phi} \geq 1/3$, for fermionic and bosonic reheating.
 
\section{Summary}
\label{summary}

Observations to date place remarkably weak constraints on the thermal history of the Universe prior to Big Bang nucleosynthesis, typically requiring only that radiation domination began at a temperature greater than about $T_{\rm RH}\sim4$~MeV \cite{tr4}. This leaves open a wide range of possible cosmological histories with low reheating temperatures, which have become a subject of much recent interest, for instance in dark matter model building. However, in the setting of low reheating temperatures, obtaining the correct baryon asymmetry via standard (thermal or non-thermal) leptogenesis and electroweak baryogenesis is generically very challenging. Specifically, for reheating temperatures $T_{\rm RH} < T_{\rm sph}\sim130$~GeV, there is a steep dilution of the baryon asymmetry after sphaleron decoupling. For the simple case of matter-like reheating, the dilution of the baryon asymmetry after sphaleron decoupling is characterized by $Y_{B} \propto \left(\frac{a_{\rm sph}}{a_{\rm RH}}\right)^3$ or equivalently $Y_{B}\propto \left(\frac{T_{\rm RH}}{T_{\rm sph}}\right)^5$. For example, for $T_{\rm RH}=1$~GeV, this dilution factor alone yields a suppression of $\mathcal{O}(10^{-11})$, making the straightforward approach to leptogenesis impossible for very low reheating temperatures. In this paper, we find the surprising result that arbitrarily low reheating temperatures above the BBN bound can be readily accommodated by simply allowing the equation of state of the oscillating inflaton condensate during reheating to be radiation-like or greater, namely $w_\phi\geq 1/3$. 

In this work, we studied leptogenesis during non-instantaneous reheating using the standard type-I seesaw framework. Our system of Boltzmann equations allowed us to track the evolution of the baryon asymmetry by simultaneously evolving the inflaton and radiation energy densities, the RHN number density, and $n_{B-L}$ throughout the reheating period. Our treatment includes both thermal and non-thermal RHN production channels; however, the non-thermal channel sourced by inflaton decays ($\phi \rightarrow NN$) is dominant throughout the parameter space we consider. We first show in Section \ref{k2} that for standard matter-like reheating ($w_\phi=0, \hspace{1mm} k=2$), it is typically impossible to obtain the correct baryon asymmetry for reheating temperatures well below $T_{\rm sph}$ due to the severe dilution penalty mentioned above. In an attempt to overcome this dilution, one may attempt to generate a large initial number of RHNs via a strong decay rate $\Gamma_{\phi \rightarrow NN}$, which would in turn lead to a large lepton asymmetry upon RHN decay. However, even in such cases, the dilution penalty combined with other constraints renders matter-like reheating incompatible with successful leptogenesis at low reheating temperatures (see, for example, the bottom right panels of Figs~\ref{k2plotv1} and \ref{k2plot}). 

The circumstances, however, are quite different when we move beyond matter-like reheating. In particular, for a generalized Starobinsky-like potential approximated by $V(\phi)\propto \phi^k$ near the minimum, the oscillating inflaton condensate during reheating has an equation of state $w_\phi = (k-2)/(k+2)$ so that $w_\phi >0$ for $k>2$. The key reason that $k>2$ is qualitatively different from matter-like reheating ($k=2$) is that the $\phi \rightarrow NN$ channel which generates the RHN population undergoes a kinematic shutoff. This is due to the fact that the decay rate for this channel is scale-factor dependent (see Eqs.~(\ref{k4boltzN}), (\ref{effma1}) and (\ref{effma2})), since it is a function of the inflaton's effective mass which is non-constant during reheating for $k\geq4$. This is ultimately due to the fact that for $k\ge 4$ the inflaton is massless in the vacuum. During reheating, the inflaton's effective mass can initially be very large, yet it decreases with increasing scale factor. As a result, an inflaton with an initial effective mass at the end of inflation $m_\phi \sim 10^{13}$ can produce a large initial population of RHNs shortly after inflation ends before its mass quickly drops below $M_N \sim 10^{12}$~GeV, at which time the $\phi \rightarrow NN$ channel shuts off. The initial density of RHNs is then essentially fixed by the inflaton-RHN coupling $y_{\phi NN}$, and the final baryon asymmetry is ultimately traced directly to this initial RHN population. 

We show that the final baryon asymmetry for $k\geq4$ is independent of $T_{\rm RH}$, up to a mild dependence on the number of relativistic degrees of freedom. Instead, the final baryon asymmetry is primarily sensitive only to $y_{\phi NN}$ and the CP-violating phase $\delta_{\rm eff}$, which is well-captured in Eqs.~(\ref{Eq:epsAnalytic}) and (\ref{deltaeff}) and Fig.~\ref{Fig:epsyPhiNN}. We find that the correct baryon asymmetry can be obtained for $10^{-9}\lesssim |\epsilon_{BL}|\lesssim 3\times10^{-4}$ and $5\times 10^{-5} \lesssim y_{\phi NN} \lesssim 10^{-2}$. The required CP-violating phase $\delta_{\rm eff}$ is inversely proportional to $y_{\phi NN}$. This can be easily understood since an increase in $y_{\phi NN}$ will generate a larger initial RHN number, thus requiring a smaller $\delta_{\rm eff}$ to produce the requisite lepton asymmetry (which is subsequently converted a baryon asymmetry via sphaleron transitions). Apart from $y_{\phi NN}$ and $|\epsilon|$, the other fundamental parameters in our model include $y_N$ and $M_N$. Interestingly, we find that the final asymmetry is effectively independent of both $y_N$ and $M_N$. For $y_N$, we derived a simple condition in Eq.~(\ref{Eq:yNconstraint}) which, when satisfied, the correct baryon asymmetry can be obtained independent from the specific value of $y_N$. For simplicity, we took $M_N=10^{12}$~GeV throughout, although a broader range of RHN masses is possible. We also note that while we used some basic MSSM parameters for concreteness, nothing about the mechanism we describe relies on supersymmetry.

In sum, we have shown that ordinary leptogenesis can be compatible with arbitrarily low reheating temperatures above the BBN limit of $T_{\rm RH}\sim 4$~MeV, by simply considering reheating scenarios where the equation of state of the oscillating inflaton condensate is $w_\phi \geq 1/3$. In these scenarios, nonthermal production of a large initial density of RHNs via inflaton decays ($\phi \rightarrow NN$) can easily lead to the requisite baryon asymmetry for a broad range of CP-violating parameters, $|\epsilon|$, in the canonical type-I seesaw framework. It is the kinematic shutoff of the nonthermal RHN production channel induced by the evolving effective mass of the inflaton which renders the $w_\phi \geq 1/3$ case categorically different from matter-like reheating with $w_\phi=0$. These results provide further support for the feasibility of low reheating temperature scenarios, which have historically been regarded as incompatible with standard baryogenesis mechanisms.

\appendix
\section{Effect of fragmentation}
\label{sec:frag}

The effects of fragmentation during reheating were discussed in detail in \cite{Garcia:2023dyf}. Here we recap the essential analytic approximations which approximate the reheating temperature as a function of the inflaton decay coupling, $y_{\phi ff}$. Note that in \cite{Garcia:2023dyf} fragmentation effects were computed numerically
assuming a T-model for inflation, rather than the Starobinsky-like model considered here. While the inflationary observables such as $n_s$ and $r$ of these are models are very similar, fragmentation is affected by the choice of model. This comes about because the normalization of the inflaton self coupling, $\lambda$ is different. For the models considered here, we have $\lambda \simeq 1.5 \times 10^{-10}$, whereas for the T-models with $k=4$, $\lambda \simeq 3.4 \times 10^{-12}$.  
Given the stronger self-coupling, we expect the effects of fragmentation to begin earlier. 

\begin{figure}[h!] \includegraphics[width=\textwidth]{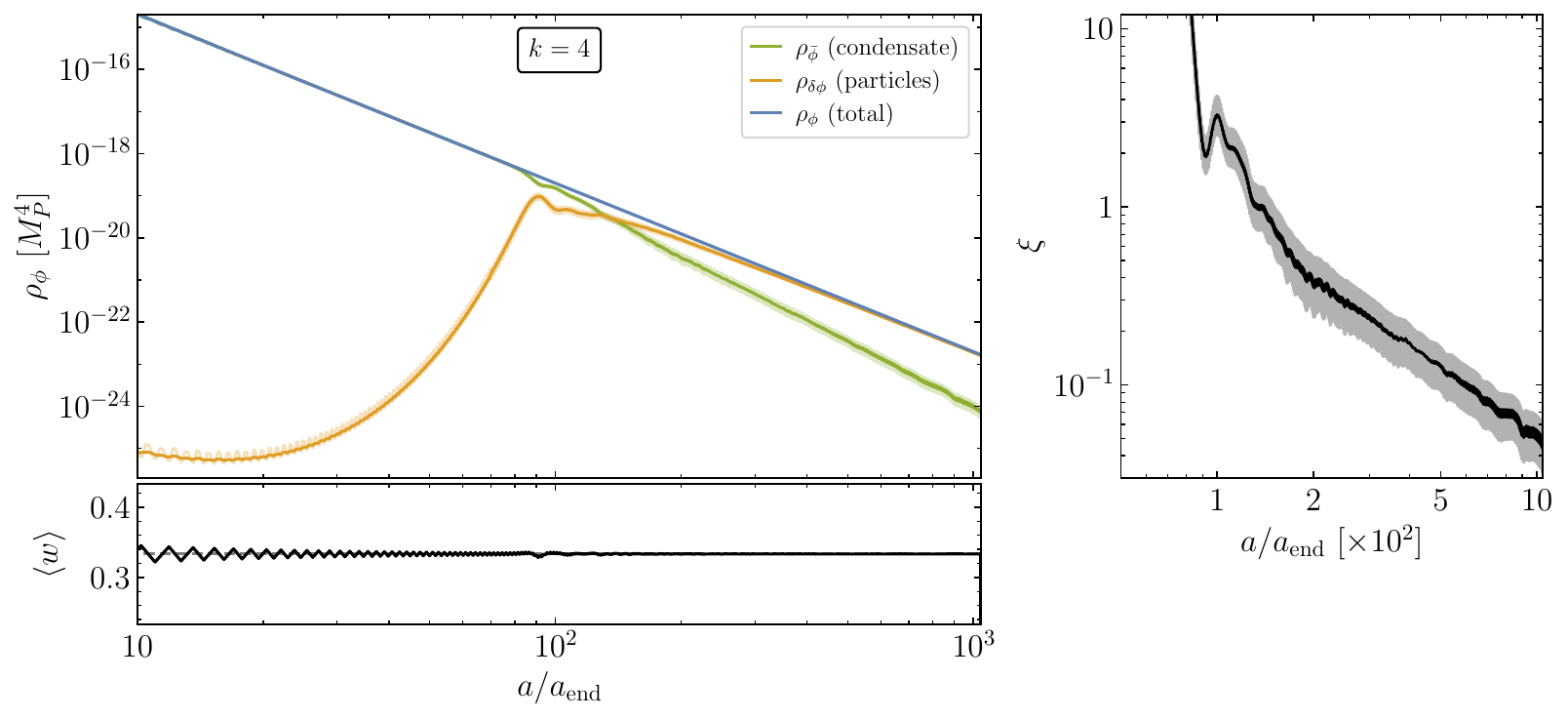} 
    \caption{Left:~inflaton energy density in the classical condensate (${\rho_{\bar \phi}}$, green), in free particles ($\rho_{\delta\phi}$, yellow) and the sum of both ($\rho_{\phi}$, blue), as functions of the scale factor. Right:~ratio of energy densities $\xi={\rho_{\bar \phi}}/\rho_{\delta\phi}$ (black) during and after fragmentation. Bottom:~the oscillation-averaged equation of state. }
    \label{frag}
\end{figure}

For the T-models with $k=4$, the onset of fragmentation occurs at $a_\beta \approx 180 \aend$. In Fig.~\ref{frag},
we show the evolution of the inflaton condensate (green) as a function of the scale factor. At $a=a_\beta \simeq 90 \aend$, the energy density of free particles (yellow) becomes significant. For $k=4$, the equation of state parameter shown in the lower panel remains approximately 1/3 throughout. On the right, we see the evolution of $\xi$ (which also differs slightly from the T-model result)
and a fit to $\xi$ for $a > a_\beta$ gives $\xi = \xi_0 (a_\beta/a)^b$ with $\xi_0 \simeq 1.1$ and $b \simeq 1.3$.

\begin{figure}[h!] \includegraphics[width=\textwidth]{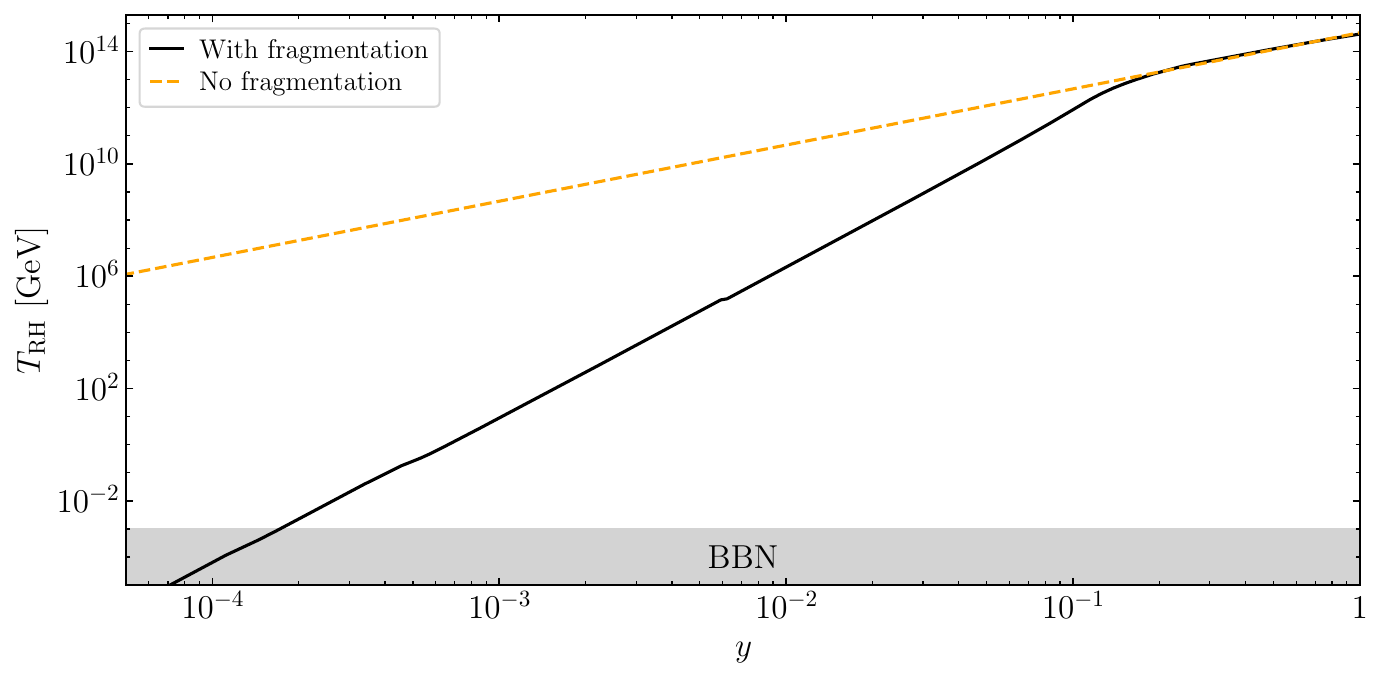} 
    \caption{The reheating temperature as a function of the inflaton coupling to fermions, $y$. The dashed line show the dependence when fragmentation is ignored. This is given by either Eq.~(\ref{trh4nofrag}) for $k= 4$. The solid curve include the effects of fragmentation. When $y$ is sufficiently large so that $\arh < a_\beta$, the solid and dotted curves track each other. At lower $y$, the reheating temperature is affected by fragmentation.}
    \label{TrehF}
\end{figure}

The fact that the energy density of the condensate does not go to zero, allows the inflaton to continue to decay after fragmentation occurs. This is because the inflaton has an effective mass which is determined by the density of the condensate. 
Thus reheating is able to complete, even if the scale factor at reheating $\arh > a_\beta$. This results in a lower reheating temperature for a given value of $y_{\phi ff}$ as can be seen by comparing Eqs.~(\ref{trh4nofrag}) and (\ref{trhr4frag}).
The numerical result for $\Trh$ as a function $y_{\phi ff}$ is shown in Fig.~\ref{TrehF}.

We can get a good analytic approximation to the reheating temperature by going back to the second of the Boltzmann equations in Eq.~(\ref{Eq:Boltz}) for radiation and keep only the source term from inflaton decay. For large enough $y_{\phi ff}$, reheating occurs before fragmentation. 
The Boltzmann equation is then
\beq
\frac{H}{a^3} \frac{d}{da}(\rho_{\rm R} a^4) = \frac43 \Gamma_{\phi} \rho_{\bar \phi} \, ,
\label{rhorfrag}
\eeq
where $\Gamma_\phi$ is given by Eq.~(\ref{k4boltzf}) and $\rho_{\bar \phi} = \rho_{\rm end} (\aend/a)^4$. This is easily integrated to give $\rho_{\rm R}$ and setting $\rho_{\rm R}(\arh) = \rho_{\bar \phi}(\arh)$ gives
\beq
\frac{\arh}{\aend}\simeq \frac{2 \sqrt{3} \pi \rho_{\rm end}^\frac12}{y_{\phi ff}^2 {\hat m} M_P} \qquad a_{\rm RH}< a_\beta \, ,
\label{arhaelt}
\eeq 
and the reheating temperature is given by 
\beq
\Trh = \frac{{{\hat m}} M_P}{2\sqrt{3} \pi \alpha^\frac14 \rho_{\rm end}^\frac14} y_{\phi ff}^2 \simeq 4.2 \times 10^{14} y_{\phi ff}^2 ~{\rm GeV} \, ,
\label{trhhighy}
\eeq
as in Eqs.~(\ref{trh4nofrag0}) and (\ref{trh4nofrag}).
This expression results in the dashed line in Fig.~\ref{TrehF}.

\begin{figure}[h!] \includegraphics[width=0.8\textwidth]{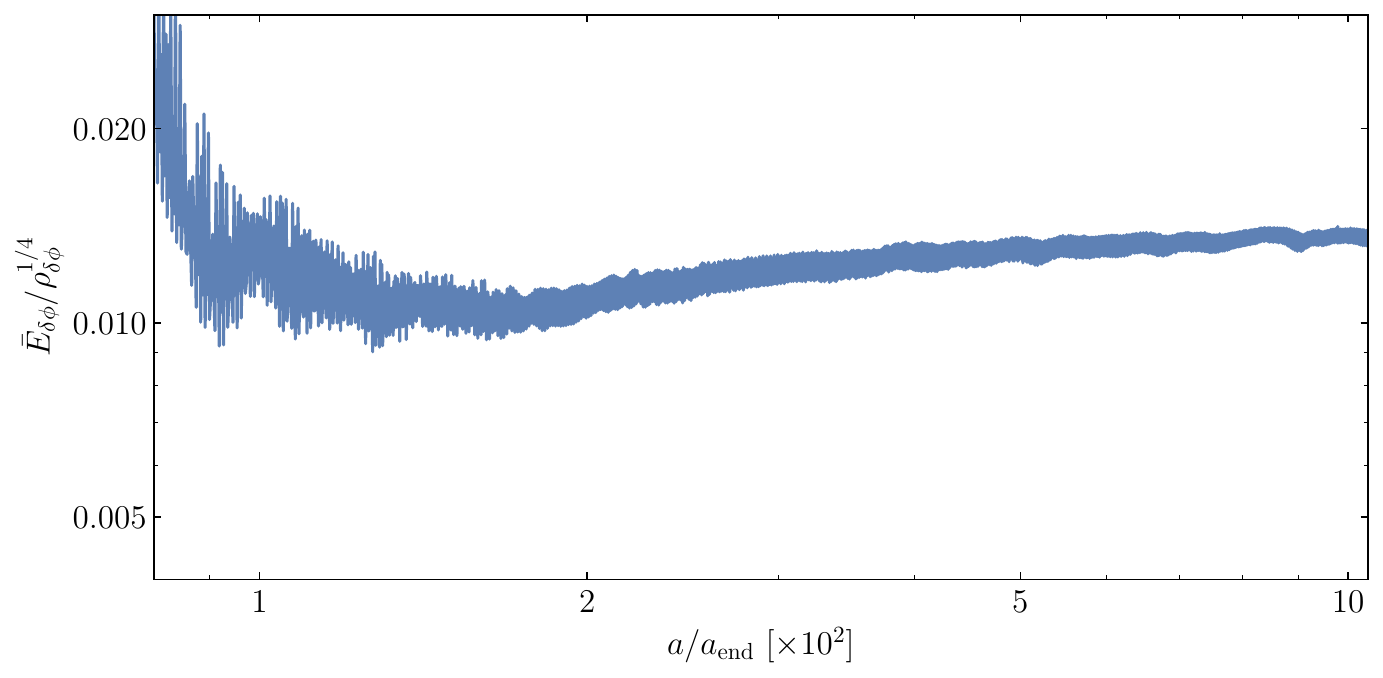} 
    \caption{Average energy per particle  }
    \label{aveE}
\end{figure}

For smaller $y_{\phi ff}$, when the effects of fragmentation are important, we must include the additional $a$-dependence in the energy density of the condensate.
The Boltzmann equation
can be rewritten as 
\beq
\frac{H}{a^3} \frac{d}{da}(\rho_{\rm R} a^4) =  \Gamma_{\delta \phi} \rho_{\delta \phi} \, ,
\label{rhorfragd}
\eeq
valid after fragmentation, so that the energy density is dominated by inflaton quanta, $\delta \phi$ with decay rate given by Eq.~(\ref{k4boltzdf}). Note that after fragmentation, there is no longer the factor of $(1+w_\phi)$ on the right hand side of the Boltzmann equation, as the decay rate pertains to free particles rather than the condensate. For $k=4$ the decay rate can be written as
\beq
\Gamma_{\delta \phi}  = c_4 \xi_0^\frac12 \beta^\frac{b}{2} \left(\frac{{\hat m}^2}{\rho_{\rm end}^\frac14} \right) \left( \frac{\aend}{a} \right)^{1+\frac{b}{2}} \, ,
\eeq
where $c_4 = y^2_{\phi ff}/8\pi c_e$,
$c_e \equiv {\bar E}/\rho_{\delta \phi}^\frac14$.
Here we will use the value of $\beta$ when the density of free inflatons dominates over the condensate. From Fig.~\ref{frag}, this occurs at
 $\beta \equiv a_{\beta}/\aend \simeq 135$.
Then Eq.~(\ref{rhorfragd}) can be solved giving
\beq
\rho_{\rm R} \simeq  2{\sqrt{3}}\frac{c_4 \xi_0^\frac12 \beta^\frac{b}{2}}{2-b} {\hat m}^2 M_P \rho_{\rm end}^\frac14  \left( \frac{\aend}{a} \right)^{3+\frac{b}{2}} \, .
\label{rhorfrag2}
\eeq
This gives us the relation between the temperature and scale factor after fragmentation, $T^4 \propto a^{3+\frac{b}{2}}$.
Finally, by setting this expression for $\rho_{\rm R}$ equal to the energy density $\rho_{\delta \phi}$ at $\arh$, we can determine $\arh$
\beq
\frac{\arh}{\aend} \simeq \left[ \frac{(2-b)}{2\sqrt{3} c_4} \xi_0^{-\frac12} \beta^\frac{-b}{2} \frac{\rho_{\rm end}^\frac34}{{\hat m}^2 M_P} \right]^\frac{2}{2-b}  \qquad a_{\rm RH} > a_\beta \, ,
\eeq
and evaluating $\rho_{\rm R}$ at $\arh$, 
we have 
\beq
\alpha_{\rm RH} \Trh^4 \simeq  \left( \frac{2 \sqrt{3} c_4 }{(2-b)}  \xi_0^\frac12 \beta^\frac{b}{2}  \frac{{\hat m}^2 M_P}{\rho_{\rm end}^\frac34}\right)^{\frac{8}{2-b}} \rho_{\rm end} \, .
\eeq
This leads to\footnote{It is important to note that this result assumes the instantaneous thermalization of the particles originating from inflaton decays. In its absence, non-perturbative effects and Pauli suppression make reheating impossible for a two-fermion final state unless $y_{\phi ff}\gtrsim (10^{10}\lambda/3)^{1/2}\sim \mathcal{O}(1)$~\cite{Bhusal:2025oqg}.}
\beq
\Trh \simeq 8.8 \times 10^{11}~{\rm GeV} \frac{y_{\phi ff}^{5.71}}{c_e^{2.86}} \simeq 2.6 \times 10^{17} ~{\rm GeV} y_{\phi ff}^{5.71}\, ,
\label{trhlowy}
\eeq
The value of $c_e$ can be determined by matching 
$\Trh$ in Eqs.~(\ref{trhhighy}) and (\ref{trhlowy})
at the value of $y_{\phi ff}\simeq 0.18$. This value of $y_{\phi ff}$ can be obtained (semi)-analytically by setting $\arh/\aend$ in Eq.~(\ref{arhaelt}) equal to $a_\beta/\aend = 135$. As one can see from Fig.~\ref{TrehF}, this is approximately the value of $y_{\phi ff}$ where the slope in $\Trh$ vs $y_{\phi ff}$ changes and the solid line separates from the dashed line. This gives $c_e \simeq 0.012$ in good agreement with the numerical result found for $c_e$ as shown in Fig.~\ref{aveE}. Thus the analytic determination of $\Trh$
 for any value of $y_{\phi ff}$ is in good agreement with our numerical results.

\section{Determining the average RHN energy for $k=2$}
\label{aveE2}

In this section, we derive the expectation value of the RHN energy to make more precise the relationship between $n_N$ and $\rho_N$. This is necessary for a consistent handling of our system of Boltzmann equations, which contains terms with both $\rho_N$ and $n_N$. In particular, because the RHNs can be relativistic or non-relativistic depending on the precise production time, the reheating history, and the temperature of the thermal bath, the equation of state associated with the RHN population is not necessarily constant throughout reheating. For instance, for $k=4$, the RHNs are relativistic immediately upon non-thermal production from inflaton decays, but they quickly become non-relativistic after kinematic shutoff of the $\phi \rightarrow NN$ channel. As a result, the RHN equation of state is not fixed throughout the evolution. For $k=2$ for the parameters we consider, the dominant channel is also the non-thermal production from inflaton decays. In this case, the channel remains kinematically open throughout the reheating period such that the dominant contributors to $\rho_N$ are relativistic; however, their momentum redshifts as $p_N \propto 1/a$. As a result, we elect to evolve $n_N$ in our Boltzmann system of equations rather than $\rho_N$. For $k=2$, we can reconstruct $\rho_N$ by determining the average energy of the RHN population $\langle E_N \rangle$. For the parameters used in this study, this average energy quickly stabilizes for $a\gg a_{\rm end}$ and remains approximately constant through the completion of reheating. 

To determine the average energy of the RHN population, we integrate the energy over all production times, weighted by the dominant nonthermal source term $S_N$ evaluated at the production time, and then divide by the total RHN number. Using the scale factor as the dynamical variable, and taking $a'$ as the scale factor at production and $a$ as the scale factor at some designated time when we want to evaluate the average RHN energy, we have 
\beq
\langle E_N \rangle (a)= \frac{\int_{a_{\rm end}}^{a} da'S_N(a')\left(\frac{a'}{a}\right)^3 E_N(a,a')}{\int_{a_{\rm end}}^{a} da' S_N(a')\left(\frac{a'}{a}\right)^3},
\label{Eq:averageRHNenergy}
\eeq with 
\beq
S_N(a')=\frac{\Gamma_{\phi NN}}{a'H(a')}\frac{\rho_\phi(a')}{m_\phi}, \hspace{5mm}\text{and } E_N(a,a')=\sqrt{M_N^2+\left(\frac{m_\phi}{2}\right)^2\left(\frac{a'}{a}\right)^2}, 
\eeq where we have used the fact that the momentum of the RHNs immediately upon decay is $\sim m_\phi/2$ for $m_\phi \gg M_N$. We can see from Eq.~\eqref{Eq:averageRHNenergy} that the factors in $S_N(a')$ which are independent of $a'$ will cancel from the numerator and denominator. Then, taking $a\gg a_{\rm end}$ and $m_\phi \gg M_N$, we find the following simple result
\begin{align}
\langle E_N \rangle (a) &= \frac{\int_{a_{\rm end}}^{a}da'a'^{3/2}(m_\phi/2)\frac{1}{a}}{\int_{a_{\rm end}}^{a}da'a'^{1/2}} \nonumber \\ & \simeq \frac{3}{5} \left(\frac{m_\phi}{2}\right).
\label{aveEN}
\end{align} 
Thus, when reconstructing $\rho_N$ for $k=2$ for plotting purposes, we use $\rho_N=\langle E_N \rangle n_N=0.6 \frac{m_\phi}{2} n_N$. We confirmed that this approximation closely agrees with the numerical result when both $\rho_N$ and $n_N$ are co-evolved and a relativistic equation of state is used for $\rho_N$. 

For $k=4$, for plotting purposes we use $\rho_N=M_N n_N$. This is well-motivated because kinematic shutoff of the $\phi \rightarrow NN$ channel occurs very early in the reheating process, namely at $a=a_*=\frac{\hat{m}}{2 M_N}$. For the generalized Starobinsky potential we consider with $k=4$, we have $\hat{m}=3.0\times 10^{13}$~GeV and we take $M_N=10^{12}$~GeV such that $a_*/a_{\rm end}=15$. For low reheating temperatures, $a_{\rm RH}$ is so large that the RHNs will be non-relativistic for essentially the entire duration of reheating. More precisely, a calculation similar to the one above can be done to determine when the expectation value of the RHN kinetic energy becomes less than say 1\% of $M_N$ (a conservative proxy for non-relativistic RHN). Doing this, we find that the kinetic energy of the RHNs will drop below 1\% of $M_N$ at about $a\simeq6a_*$. Recalling that for $k=4$, $a_{\rm RH}/a_{\rm end}=\mathcal{O}(10^{15})$ for $T_{\rm RH}=1$~GeV, we see that non-relativistic RHN is a very good approximation.

\acknowledgments
 This work was supported in part by DOE grant DE-SC0011842 at the University of Minnesota. The work of M.A.G.G. was supported by the DGAPA-PAPIIT grant IA100525 at UNAM, and a Cátedra Marcos Moshinsky.


\begin{thebibliography}{99}


  \bibitem{CFOY}
R.~H.~Cyburt, B.~D.~Fields, K.~A.~Olive and T.-H. Yeh,
  Rev.\ Mod.\ Phys.\  {\bf 88}, 015004 (2016)
  [arXiv:1505.01076 [astro-ph.CO]].


\bibitem{coc18}
   C.~Pitrou, A.~Coc, J.~P.~Uzan and E.~Vangioni,
  Phys.\ Rept.\  {\bf 754}, 1 (2018)
  [arXiv:1801.08023 [astro-ph.CO]].

\bibitem{foyy}
B.~D.~Fields, K.~A.~Olive, T.~H.~Yeh and C.~Young,
JCAP \textbf{03}, 010 (2020)
[erratum: JCAP \textbf{11}, E02 (2020)]
[arXiv:1912.01132 [astro-ph.CO]].

 

\bibitem{Kolb:1983ni}
E.~W.~Kolb and M.~S.~Turner,
Ann. Rev. Nucl. Part. Sci. \textbf{33}, 645-696 (1983)

\bibitem{Fukugita:1986hr}
M.~Fukugita and T.~Yanagida,
Phys. Lett. B \textbf{174}, 45-47 (1986)

\bibitem{Olive:1994xw}
K.~A.~Olive,
Lect. Notes Phys. \textbf{440}, 1-37 (1994)
[arXiv:hep-ph/9404352 [hep-ph]].

\bibitem{Riotto:1999yt}
A.~Riotto and M.~Trodden,
Ann. Rev. Nucl. Part. Sci. \textbf{49}, 35-75 (1999)
[arXiv:hep-ph/9901362 [hep-ph]].

\bibitem{Bodeker:2020ghk}
D.~Bodeker and W.~Buchmuller,
Rev. Mod. Phys. \textbf{93}, no.3, 3 (2021)
[arXiv:2009.07294 [hep-ph]].

\bibitem{Staro}
A.~A.~Starobinsky,
Phys.\ Lett.\ B {\bf 91}, 99 (1980).

\bibitem{inflation}
A.~H.~Guth,
  Phys.\ Rev.\ D {\bf 23} (1981) 347.


 

\bibitem{reviews}
   K.~A.~Olive,
  Phys.\ Rept.\  {\bf 190} (1990) 307;
A. D. Linde, {\it Particle  
Physics and
Inflationary Cosmology} (Harwood, Chur, Switzerland, 1990); 
  D.~H.~Lyth and A.~Riotto,
{\it Phys.\ Rep.}  {\bf 314} (1999) 1
[arXiv:hep-ph/9807278];
J.~Martin, C.~Ringeval and V.~Vennin,
  Phys.\ Dark Univ.\  {\bf 5-6}, 75-235 (2014)
  [arXiv:1303.3787 [astro-ph.CO]];
  J.~Martin, C.~Ringeval, R.~Trotta and V.~Vennin,
  JCAP {\bf 1403} (2014) 039
  [arXiv:1312.3529 [astro-ph.CO]];
 J.~Martin,
  Astrophys.\ Space Sci.\ Proc.\  {\bf 45}, 41 (2016)
  [arXiv:1502.05733 [astro-ph.CO]];
  J.~Ellis and D.~Wands,
arXiv:2312.13238 [astro-ph.CO], in
S.~Navas \textit{et al.} [Particle Data Group],
Phys. Rev. D \textbf{110} (2024) no.3, 030001

\bibitem{tr4}
M.~Kawasaki, K.~Kohri and N.~Sugiyama,
Phys. Rev. D \textbf{62}, 023506 (2000)
[arXiv:astro-ph/0002127 [astro-ph]].
P.~F.~de Salas, M.~Lattanzi, G.~Mangano, G.~Miele, S.~Pastor and O.~Pisanti,
Phys. Rev. D \textbf{92}, no.12, 123534 (2015)
[arXiv:1511.00672 [astro-ph.CO]].
S.~Hannestad,
Phys. Rev. D \textbf{70}, 043506 (2004)
[arXiv:astro-ph/0403291 [astro-ph]];
T.~Hasegawa, N.~Hiroshima, K.~Kohri, R.~S.~L.~Hansen, T.~Tram and S.~Hannestad,
JCAP \textbf{12}, 012 (2019)
[arXiv:1908.10189 [hep-ph]].

\bibitem{Bhattiprolu:2022sdd}
P.~N.~Bhattiprolu, G.~Elor, R.~McGehee and A.~Pierce,
JHEP \textbf{01}, 128 (2023)
[arXiv:2210.15653 [hep-ph]].

\bibitem{Cosme:2023xpa}
C.~Cosme, F.~Costa and O.~Lebedev,
Phys. Rev. D \textbf{109}, no.7, 075038 (2024)
[arXiv:2306.13061 [hep-ph]].

\bibitem{Silva-Malpartida:2023yks}
J.~Silva-Malpartida, N.~Bernal, J.~Jones-P{\'e}rez and R.~A.~Lineros,
JCAP \textbf{09}, 015 (2023)
[arXiv:2306.14943 [hep-ph]].

\bibitem{Arcadi:2024wwg}
G.~Arcadi, F.~Costa, A.~Goudelis and O.~Lebedev,
JHEP \textbf{07}, 044 (2024)
[arXiv:2405.03760 [hep-ph]].

\bibitem{Boddy:2024vgt}
K.~K.~Boddy, K.~Freese, G.~Montefalcone and B.~Shams Es Haghi,
Phys. Rev. D \textbf{111}, no.6, 6 (2025)
[arXiv:2405.06226 [hep-ph]].

\bibitem{Belanger:2024yoj}
G.~B{\'e}langer, N.~Bernal and A.~Pukhov,
JHEP \textbf{03}, 079 (2025)
[arXiv:2412.12303 [hep-ph]].

\bibitem{Amiri:2025ras}
A.~Amiri, B.~Diaz Saez and K.~M{\"o}hling,
[arXiv:2511.21520 [hep-ph]].

\bibitem{Henrich:2025pca}
S.~E.~Henrich, Y.~Mambrini and K.~A.~Olive,
JCAP \textbf{04}, 068 (2026)
[arXiv:2512.04229 [hep-ph]].

\bibitem{Henrich:2026tox}
S.~E.~Henrich, Y.~Mambrini and K.~A.~Olive,
[arXiv:2605.03014 [hep-ph]].


\bibitem{Kuzmin:1985mm}
V.~A.~Kuzmin, V.~A.~Rubakov and M.~E.~Shaposhnikov,
Phys. Lett. B \textbf{155}, 36 (1985)


\bibitem{Arnold:1987mh}
P.~B.~Arnold and L.~D.~McLerran,
Phys. Rev. D \textbf{36}, 581 (1987);
P.~B.~Arnold and L.~D.~McLerran,
Phys. Rev. D \textbf{37}, 1020 (1988)

\bibitem{spha2}
  S.~Y.~Khlebnikov and M.~E.~Shaposhnikov,
Nucl.\ Phys.\ B {\bf 308}, 885 (1988).


 \bibitem{seesaw}
P.~Minkowski,
  Phys.\ Lett.\ B {\bf 67} (1977) 421;
M.~Gell-Mann, P.~Ramond and R.~Slansky, in {\it Supergravity}, eds. D. Freedman and P. Van Nieuwenhuizen
(North Holland, Amsterdam, 1979), pp. 315-321. ISBN 044485438x;
T. Yanagida, in {\it Proceedings of the Workshop on 
the Unified Theory and The Baryon Number of the Universe}, eds O. Sawada  
and S. Sugamoto. KEK79-18 (1979);
R.~N.~Mohapatra and G.~Senjanovic,
  Phys.\ Rev.\ Lett.\  {\bf 44}, 912 (1980);
J.~Schechter and J.~W.~F.~Valle,
  Phys.\ Rev.\ D {\bf 22} (1980) 2227;
J.~Schechter and J.~W.~F.~Valle,
  Phys.\ Rev.\ D {\bf 25} (1982) 774.




\bibitem{Buchmuller:2002rq}
W.~Buchmuller, P.~Di Bari and M.~Plumacher,
Nucl. Phys. B \textbf{643}, 367-390 (2002)
[erratum: Nucl. Phys. B \textbf{793}, 362 (2008)]
[arXiv:hep-ph/0205349 [hep-ph]].

\bibitem{Buchmuller:2003gz}
W.~Buchmuller, P.~Di Bari and M.~Plumacher,
Nucl. Phys. B \textbf{665}, 445-468 (2003)
[arXiv:hep-ph/0302092 [hep-ph]].

\bibitem{Chankowski:2003rr}
P.~H.~Chankowski and K.~Turzynski,
Phys. Lett. B \textbf{570}, 198-204 (2003)
[arXiv:hep-ph/0306059 [hep-ph]].

 

\bibitem{Giudice:2003jh}
G.~F.~Giudice, A.~Notari, M.~Raidal, A.~Riotto and A.~Strumia,
Nucl. Phys. B \textbf{685}, 89-149 (2004)
[arXiv:hep-ph/0310123 [hep-ph]].

 \bibitem{DIbound}
 S.~Davidson and A.~Ibarra,
  Phys.\ Lett.\ B {\bf 535} (2002) 25
  [hep-ph/0202239];

\bibitem{Lazarides:1990huy}
G.~Lazarides and Q.~Shafi,
Phys. Lett. B \textbf{258}, 305-309 (1991)

\bibitem{Campbell:1992hd}
B.~A.~Campbell, S.~Davidson and K.~A.~Olive,
Nucl. Phys. B \textbf{399} (1993), 111-136
[arXiv:hep-ph/9302223 [hep-ph]].

\bibitem{Giudice:1999fb}
G.~F.~Giudice, M.~Peloso, A.~Riotto and I.~Tkachev,
JHEP \textbf{08}, 014 (1999)
[arXiv:hep-ph/9905242 [hep-ph]].

\bibitem{Asaka:1999yd}
T.~Asaka, K.~Hamaguchi, M.~Kawasaki and T.~Yanagida,
Phys. Lett. B \textbf{464} (1999), 12-18
[arXiv:hep-ph/9906366 [hep-ph]].

\bibitem{Hahn-Woernle:2008tsk}
F.~Hahn-Woernle and M.~Plumacher,
Nucl. Phys. B \textbf{806}, 68-83 (2009)
[arXiv:0801.3972 [hep-ph]].

\bibitem{Kanemura:2025rct}
S.~Kanemura, K.~Kaneta and D.~Nanda,
Phys. Rev. D \textbf{113}, no.5, 055046 (2026)
[arXiv:2508.00315 [hep-ph]].

\bibitem{DOnofrio:2014rug}
M.~D'Onofrio, K.~Rummukainen and A.~Tranberg,
Phys. Rev. Lett. \textbf{113}, no.14, 141602 (2014)
[arXiv:1404.3565 [hep-ph]].

\bibitem{Hamada:2015xva}
Y.~Hamada and K.~Kawana,
Phys. Lett. B \textbf{763}, 388-392 (2016)
[arXiv:1510.05186 [hep-ph]].

\bibitem{Zhang:2023oyo}
X.~Zhang,
JHEP \textbf{05}, 147 (2024)
[arXiv:2311.05824 [hep-ph]].

\bibitem{Barman:2024ujh}
B.~Barman, A.~Basu, D.~Borah, A.~Chakraborty and R.~Roshan,
Phys. Rev. D \textbf{111}, no.5, 055016 (2025)
[arXiv:2410.19048 [hep-ph]].

\bibitem{GKMO2}
M.~A.~G.~Garcia, K.~Kaneta, Y.~Mambrini and K.~A.~Olive,
JCAP \textbf{04}, 012 (2021)
[arXiv:2012.10756 [hep-ph]].

\bibitem{Ellis:2021kad}
J.~Ellis, M.~A.~G.~Garcia, D.~V.~Nanopoulos, K.~A.~Olive and S.~Verner,
Phys. Rev. D \textbf{105}, no.4, 043504 (2022)
[arXiv:2112.04466 [hep-ph]].

\bibitem{Planck}
N.~Aghanim \textit{et al.} [Planck],
Astron. Astrophys. \textbf{641}, A6 (2020)
[erratum: Astron. Astrophys. \textbf{652}, C4 (2021)]
[arXiv:1807.06209 [astro-ph.CO]];
Y.~Akrami \textit{et al.} [Planck],
Astron. Astrophys. \textbf{641}, A10 (2020)
[arXiv:1807.06211 [astro-ph.CO]].




\bibitem{LiddleLeach} 
  A.~R.~Liddle and S.~M.~Leach,
  Phys.\ Rev.\ D {\bf 68}, 103503 (2003)
  [astro-ph/0305263];
  
\bibitem{Martin:2010kz}
J.~Martin and C.~Ringeval,
Phys. Rev. D \textbf{82}, 023511 (2010)
[arXiv:1004.5525 [astro-ph.CO]].

\bibitem{Ellis:2025zrf}
J.~Ellis, M.~A.~G.~Garcia, K.~A.~Olive and S.~Verner,
Phys. Rev. D \textbf{113}, no.6, 063571 (2026)
[arXiv:2510.18656 [hep-ph]].

\bibitem{Giudice:2000ex}
G.~F.~Giudice, E.~W.~Kolb and A.~Riotto,
Phys. Rev. D \textbf{64}, 023508 (2001)
[arXiv:hep-ph/0005123 [hep-ph]];
 D.~J.~H.~Chung, E.~W.~Kolb and A.~Riotto,
  Phys.\ Rev.\ D {\bf 60} (1999) 063504
  [hep-ph/9809453].

\bibitem{Ellis:2015jpg}
J.~Ellis, M.~A.~G.~Garc{\' i}a, D.~V.~Nanopoulos, K.~A.~Olive and M.~Peloso,
JCAP \textbf{03}, 008 (2016)
[arXiv:1512.05701 [astro-ph.CO]].

\bibitem{Clery:2021bwz}
S.~Clery, Y.~Mambrini, K.~A.~Olive and S.~Verner,
Phys. Rev. D \textbf{105}, no.7, 075005 (2022)
[arXiv:2112.15214 [hep-ph]].

\bibitem{Scherrer:1984fd}
R.~J.~Scherrer and M.~S.~Turner,
Phys. Rev. D \textbf{31}, 681 (1985)

  \bibitem{GKMO1}
M.~A.~G.~Garcia, K.~Kaneta, Y.~Mambrini and K.~A.~Olive,
Phys. Rev. D \textbf{101} (2020) no.12, 123507
[arXiv:2004.08404 [hep-ph].


\bibitem{HT}
  J.~A.~Harvey and M.~S.~Turner,
Phys.\ Rev.\ D {\bf 42}, 3344 (1990).

\bibitem{Yeh:2026pil}
T.~H.~Yeh, K.~A.~Olive, B.~D.~Fields, E.~Aver, R.~W.~Pogge, N.~S.~J.~Rogers, E.~D.~Skillman and M.~K.~Weller,
[arXiv:2601.22239 [astro-ph.CO]].


\bibitem{Davidson:2000dw}
S.~Davidson, M.~Losada and A.~Riotto,
Phys. Rev. Lett. \textbf{84}, 4284-4287 (2000)
[arXiv:hep-ph/0001301 [hep-ph]].

\bibitem{kmov}
K.~Kaneta, Y.~Mambrini, K.~A.~Olive and S.~Verner,
Phys. Rev. D \textbf{101}, no.1, 015002 (2020)
[arXiv:1911.02463 [hep-ph]].





  \bibitem{luty}
  M.~A.~Luty,
  Phys.\ Rev.\ D {\bf 45}, 455 (1992).


\bibitem{CPviol}
  L.~Covi, E.~Roulet and F.~Vissani,
Phys.\ Lett.\ B {\bf 384}, 169 (1996).
[hep-ph/9605319].
  M.~Flanz, E.~A.~Paschos and U.~Sarkar,
Phys.\ Lett.\ B {\bf 345}, 248 (1995), Erratum: [Phys.\ Lett.\ B {\bf 384}, 487 (1996)], Erratum: [Phys.\ Lett.\ B {\bf 382}, 447 (1996)].
[hep-ph/9411366].

\bibitem{Campbell:1990fa}
B.~A.~Campbell, S.~Davidson, J.~R.~Ellis and K.~A.~Olive,
Phys. Lett. B \textbf{256}, 484-490 (1991)

\bibitem{Campbell:1991at}
B.~A.~Campbell, S.~Davidson, J.~R.~Ellis and K.~A.~Olive,
Astropart. Phys. \textbf{1}, 77-98 (1992)

\bibitem{Fischler:1990gn}
W.~Fischler, G.~F.~Giudice, R.~G.~Leigh and S.~Paban,
Phys. Lett. B \textbf{258}, 45-48 (1991)

\bibitem{Ibanez:1992aj}
L.~E.~Ibanez and F.~Quevedo,
Phys. Lett. B \textbf{283}, 261-269 (1992)
[arXiv:hep-ph/9204205 [hep-ph]].

\bibitem{Garcia:2023eol}
M.~A.~G.~Garcia and M.~Pierre,
JCAP \textbf{11} (2023), 004
[arXiv:2306.08038 [hep-ph]].


\bibitem{Garcia:2023dyf}
M.~A.~G.~Garcia, M.~Gross, Y.~Mambrini, K.~A.~Olive, M.~Pierre and J.~H.~Yoon,
JCAP \textbf{12}, 028 (2023)
[arXiv:2308.16231 [hep-ph]].

\bibitem{Clery:2024dlk}
S.~Clery, M.~A.~G.~Garcia, Y.~Mambrini and K.~A.~Olive,
Phys. Rev. D \textbf{109}, no.10, 103540 (2024)
[arXiv:2402.16958 [hep-ph]].

\bibitem{Cosme:2024ndc}
C.~Cosme, F.~Costa and O.~Lebedev,
JCAP \textbf{06}, 031 (2024)
[arXiv:2402.04743 [hep-ph]].

\bibitem{Co:2022bgh}
R.~T.~Co, Y.~Mambrini and K.~A.~Olive,
Phys. Rev. D \textbf{106}, no.7, 075006 (2022)
[arXiv:2205.01689 [hep-ph]].

\bibitem{Barman:2022qgt}
B.~Barman, S.~Cl{\'e}ry, R.~T.~Co, Y.~Mambrini and K.~A.~Olive,
JHEP \textbf{12}, 072 (2022)
[arXiv:2210.05716 [hep-ph]].

\bibitem{Clery:2022wib}
S.~Clery, Y.~Mambrini, K.~A.~Olive, A.~Shkerin and S.~Verner,
Phys. Rev. D \textbf{105}, no.9, 095042 (2022)
[arXiv:2203.02004 [hep-ph]].

\bibitem{Bhusal:2025oqg}
N.~Bhusal, M.~E.~C.~M., M.~A.~G.~Garcia, A.~G.~Menkara and M.~Pierre,
JCAP \textbf{06}, 064 (2026)
[arXiv:2512.16203 [hep-ph]].


\end{thebibliography}

\end{document}